\documentclass[english,natbib]{svjour3}
\usepackage[T1]{fontenc}
\usepackage[latin9]{inputenc}
\usepackage{textcomp}
\usepackage{amsmath}
\usepackage{graphicx}
\usepackage{babel}
\usepackage{color}
\usepackage[none]{hyphenat}

\begin{document}

\title{Non-Ideality of a DNA Strand Displacement AND Gate Studied with a Dynamic Bonded DNA Model}

\titlerunning{Non-ideality of a DNA strand displacement AND gate}

\author{Carsten Svaneborg \and Harold Fellermann \and Steen Rasmussen}

\authorrunning{Svaneborg, Fellermann and Rasmussen}

\institute
{
    C. Svaneborg \and H. Fellermann \and S. Rasmussen    
    \at 
   Centre for Fundamental Living Technology, Department of Physics, Chemistry
and Pharmacy, University of Southern Denmark, Campusvej 55, DK-5320 Odense,
Denmark, \email{science@zqex.dk}, \email{harold@sdu.dk}, and \email{steen@sdu.dk}
\and
   H. Fellermann \at
Complex Systems Lab, Barcelona Biomedical Research Park, Universitat Pompeu
Fabra, Dr. Aiguad\'e 88, 08003 Barcelona, Spain,
\and
   S. Rasmussen \at
Santa Fe Institute, 1399 Hyde Park Road, Santa Fe NM 87501, USA
}

\date{Received: date / Accepted: date}

\maketitle
\begin{abstract}
We perform a spatially resolved simulation study of an AND gate based on DNA
strand displacement using several lengths of the toehold and the adjacent domains.
DNA strands are modelled using a coarse-grained dynamic bonding model {[}C. Svaneborg, Comp. Phys. Comm. 183, 1793 (2012){]}.
We observe a complex transition path from the initial state to the final state of
the AND gate. This path is strongly influenced by non-ideal effects due to transient
bubbles revealing undesired toeholds and thermal melting of whole strands.
We have also characterised the bound and unbound kinetics of single strands¸ and
in particular the kinetics of the total AND operation and the three distinct
distinct DNA transitions that it is based on.
We observe a exponential kinetic dependence on the toehold length of the
competitive displacement operation, but that the gate operation time is
only weakly dependent on both the toehold and adjacent domain length. Our gate
displays excellent logical fidelity in three input states, and quite poor
fidelity in the fourth input state. This illustrates how non-ideality can have
very selective effects on fidelity. Simulations and detailed analysis such as
those presented here provide molecular insights into strand displacement
computation, that can be also be expected in chemical implementations.


\end{abstract}

\section{Introduction}

Ever since the pioneering work of Adlemann in 1994~\citep{Adl:1994},
DNA has been recognized as a massively parallel, versatile, and inexpensive
computing substrate. In order for such substrate to be of practical
interest, however, it is desirable that the computational framework
is scalable and that individual computational elements can be combined to
form more complex circuits. Recently, a scalable approach to enzyme-free DNA
computing has been proposed where logic gates consist of relatively short DNA
strands that communicate via strand displacement~\citep{See:06,Qia:2011}.
Strand displacement is also used to control the communication between DNA decorated 
chemical containers~\citep{Maik1:2010,hadorn2012specific,Benny:2013,amos2011biological}.

In the strand displacement approach to DNA computation, individual gates
consist of one DNA template that is composed of several logical domains.
In their initial state, all
domains but one are hybridized to one or more complementary strands
and are therefore inert. The only exposed single strand domain of
each gate is a short toehold region at one end of the template. This
toehold region can reversibly bind a complementary signal strand which
is designed to be longer than the toehold domain and complementary
to the next domain(s) of the template. The newly binding signal is
then able to hybridize to all matching domains of the template, thereby
displacing strands that where previously bound and possibly exposing new
toeholds~\citep{Zha:2009}. The displaced strands can be fluorescent
output signals, or internal signals that can bind to toehold
regions of downstream gates. Ideally, by choosing domains and toeholds
of appropriate length, toehold binding will be reversible whereas
the total strand displacement process is irreversible. 
Hence computation is energetically downhill
and kinetically irreversible, if and only if the correct input strands
are present and match the logical setup of the gates.  It has been shown
that this approach leads to modular logic gates that enable the design
of large scale DNA circuits~\citep{Car:2011,Lak:2012}.

\begin{figure}
\centering
\includegraphics[width=0.99\textwidth]{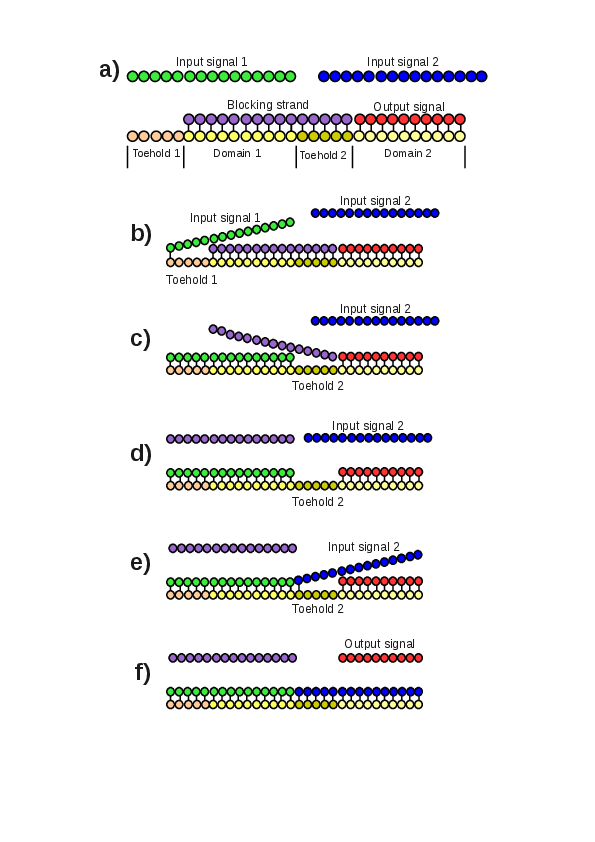}
\caption{Design of the AND gate using a template with two toeholds and domains.
Ideally the gate operates as follows: a) initially toehold 1 is exposed, a
blocking strand (magenta) protects toehold 2, and the output signal strand (red)
is hybridized with domain 2. b) Signal strand 1 (green) hybridizes with toehold
1 and c) reversibly displaces the blocking strand whereby d) toehold 2 is freed.
e) Signal strand 2 (blue) hybridizes with toehold 2 and f) irreversibly
displaces the output signal strand (red).
}
\protect\label{fig:and-model}
\end{figure}


Despite the fact that DNA self-assembly and strand-displacement operations are widely
utilized in the field of DNA nanotechnology, only little is known about their
kinetics~\citep{Zha:2009}. In the present paper,
we study effects of non-ideality on the kinetics and fidelity of a strand displacement AND gate using a spatially resolved
coarse-grained DNA model.~\citep{SvaneborgCPC2012} Fig. \ref{fig:and-model} shows the
AND gate design that we are simulating. It consists of two toeholds of identical
length $t$ and two adjacent domains of identical length $l$. Note that we do not
presently worry about garbage collecting the blocking strand after it has been
displaced \citep{Car:2011}.

When DNA computation is implemented in terms of real molecular reactions
we have to content with non-ideality. The most important source of non-ideality
is that experiments are carried out at non-zero temperature. The probability of
a given molecular DNA state is not just determined by the binding energy (as
at absolute zero temperature), but by the free energy which also contains entropic
contributions. The most likely molecular state is that which has a minimal free energy,
which is a temperature dependent balance between minimizing the energy and
maximizing the entropy. At finite temperatures, stretches of DNA nucleotides
will spontaneously dehybridize, such that the resulting bubbles increase the
configurational entropy and hence reduce the free energy of the double strand~\citep{PhysRevLett.90.138101,PhysRevE.68.061911,jost2009unified}. 
As the temperature is increased towards the melting temperature (where half of the
hybridization bonds are present on average), these bubbles grow in size until
the double strand thermally melts forming two free strands. 
Bubbles are more likely to be found at the end of a double stranded domain,
as these are less stable due to the lacking stacking interaction at the end~\citep{SantaLucia}.
When two neighbouring strands are hybridized to a template, a single backbone
bond is also missing at the interface between them. And this nick is also less
stable~\citep{Lane01021997,protozanova2004stacked}.
These effects of non-idealty affect DNA strand computation operation since
bubbles transiently expose nucleotides and hence enables undesired strand
displacement operations to take place. Thermal melting can also release
strands that were expected to be irreversible bound to the template. 
Transient states along pathway from an initial to a final states can also
have a higher free energy. In this case, the transition state with the highest
free energy along the pathway will determine the transition rate and it will
depend exponentially on the height of the free energy barrier relative to
the initial state~\citep{kramers1940}. For instance, it has been observed that
strand displacement slows down exponentially as the toeholds become longer.~\citep{Yurke:2003}.

The chemical structure of short DNA oligomers can be studied with atomistic
molecular dynamics simulations such as Amber~\citep{case2010amber,cheatham2000molecular}
and Charmm~\citep{CHARMM2009,mackerell2000development}. However, if we are
only interested in secondary structure, it is more effective to utilize
coarse-grained simulation models. Coarse-graining is a statistical physical
technique by which irrelevant microscopic details are systematically
removed, producing an effective model~\citep {langowski2006polymer,de2011polymer}
with the same mesoscopic properties. The major computational advantage of
coarse-graining is that it allows us to focus our computational resources
on studying the mesoscopic DNA structures and dynamics of interest. 

Coarse-grained models describe a nucleotide by a small number of effective interaction sites.
In the ``three sites per nucleotide'' model of de Pablo and co-workers, three sites represent
the phosphate backbone site, the sugar group, and the base, respectively~\citep{Sambriski2009,sambriski2009sequence}.
There is also a number of ``two sites per nucleotide'' models, e.g. the model of Ouldridge
and co-workers~\citep{ouldridge2010dna,OulridgeLouisDoye2011}, where one site represents the
base and another site the backbone and the sugar ring. Savelyev and Papoian~\citep{SavelyevPapoian2011}
have formulated a ``one site per nucleotide'' model. As the number of interaction sites per
nucleotide is reduced, the chemical structure is progressively lost. In simulations of DNA
tagged nano-particles, even more coarse-grained models are used. DNA molecules have been
modelled e.g. as semi-flexible polymers with attractive sites on each monomer~\citep{hsu2010theoretical},
or as a single sticky site that can be hybridized with free complementary free sticky sites~\citep{Martines-VeracoecheaPRL2011}.
While the chemical structure of DNA has been completely eliminated, these models still retain
the DNA sequence specific hybridization effects on nano-particle self-assembly.

We are interested in studying the statistical mechanics of hybridizing DNA strands
and in particular the kinetics of DNA self-assembly and DNA computation using a DNA
model that is as coarse-grained as possible. We have implemented a general framework
that allows for directional bonds to be reversibly formed and broken during molecular
dynamics simulations~\citep{SvaneborgCPC2012}. Along with the bonds, the angular and
dihedral interactions required to model the residual effects of chemical structure
are also dynamically introduced and removed as dictated by the bond dynamics. This
framework allows us to simulate reversible hybridization of complementary beads and
chains built from such beads. In the present paper, we study a minimal dynamic bonding
DNA model. For simplicity, we assume that the binding energy, as well as the bond, angular, and dihedral
potentials are independent of sequence, and we have chosen a force field that produces 
a flat ladder-like structure in the double stranded state. Our motivation for these
choices is to minimize the number of required model parameters.

Dynamic bonding DNA models combine ideas from most of the existing DNA models.
We regard them as dynamic generalizations of statistical mechanical theories
and simultaneously as simplifications of coarse-grained DNA models.
As in the Poland-Scheraga model of DNA melting~\citep{poland1966phase,jost2009unified}, 
 complementary base pairs can either be hybridized or open. When a base pair is hybridized, it is characterized by a continuous hybridization potential as in the Dauxois-Peyrard-Bishop model~\citep{peyrard2008modelling}. Dynamic bonding DNA models can also be regarded as off-lattice
 generalizations of the lattice Poland-Scheraga model \citep{EveraersKumarSimm2007}.
 Rather than trying to model chemical structure with interaction sites as in the
 ``two and three sites per nucleotide'' models~\citep{Sambriski2009,sambriski2009sequence,ouldridge2010dna,OulridgeLouisDoye2011} 
dynamic bonding DNA models use angular and dihedral interactions to model the residual
effects of local chemical structure. Dynamic bonded DNA double stands can reversibly
melt and re-anneal, which is not possible with the ``one site per nucleotide model''
of Savelyev and Papoian~\citep{SavelyevPapoian2011} due to the special ``fan'' interactions
it uses between stretches of opposing nucleotides. Finally, as in the sticky DNA models
\citep{Martines-VeracoecheaPRL2011}, a single bead in a dynamic bonding DNA model can
equally well represent a domain. 

Sect.~\ref{sec:DNA-model} introduces the dynamic bonding DNA model. We present the
simulation results and discussion in Sect.~\ref{sec:Results} and a conclusion in
Sect.~\ref{sec:Conclusions}.

\section{Model\label{sec:DNA-model}}

In the present dynamic bonding DNA model, single stranded DNA (ssDNA) is represented
by a string of nucleotide beads connected by stiff springs representing directional
backbone bonds. Complementary beads can reversibly form hybridization bonds. Rather
than limiting the model to represent the ACGT nucleotides, we increase the alphabet
maximally to avoid getting trapped in misaligned transient hybridization states.
A novel feature of our DNA model is that it involves dynamic hybridization
bonds, which are introduced or removed between complementary interaction
sites or beads when they enter or exit the hybridization reaction radius.
Along with the bonds, we dynamically introduce or remove angular and dihedral
interactions in the chemical neighbourhood of a hybridizing bead pair.
These interactions are introduced based on the local bond and bead type
pattern, and hence allows us to retain some effects of the local chemical
structure in coarse-grained models. We utilize bonds carrying directionality
to represent the 3'-5' backbone structure of DNA molecules. This allows us
to introduce dihedral interactions that can distinguish between parallel
and anti-parallel strand alignments. We have implemented this framework
in a modified version of the Large-scale Atomic/Molecular Massively Parallel
Simulator (LAMMPS)~\citep{Lammps,SvaneborgCPC2012}. 

The DNA model relies on two ingredients, a Langevin dynamic for propagating
a system in time and space, and a dynamic directional bonding scheme~\citep{SvaneborgCPC2012}
that propagates the chemical structure of the system. The force on bead $i$
is given by a Langevin equation

\[
{\bf F}_{i}=-\boldsymbol\nabla_{{\bf R}_{i}}U-\frac{m}{\Gamma}{\dot{\bf R}_{i}}+{\boldsymbol\xi}_{i}\quad\text{with}\quad U=U_\text{bond}+U_\text{angle}+U_\text{dihedral}+U_\text{pair}.
\]

Here, the first term denotes a conservative force derived from the
potential $U$. The second term is a velocity dependent friction,
and the third a stochastic driving force characterized by
$\langle{\boldsymbol\xi}_{i}(t)\rangle=0$ and 
$\langle{\boldsymbol\xi}_{i}(t){\boldsymbol\xi}_{j}(t')\rangle=k_{B}Tm/(\Gamma\Delta t)\delta_{ij}\delta(t-t')$.
The potential $U$ comprises four terms representing bond, angular, dihedral, and
non-bonded pair interactions, respectively. The friction and stochastic
driving force implicitly represents the effect of a solvent with a specified
friction and temperature. The Langevin dynamics is integrated using a
Velocity Verlet algorithm with a time step $\Delta t=0.001\tau_L$ and
$\Gamma=2\tau_L$ using a customized version of LAMMPS~\citep{Lammps,SvaneborgCPC2012}.

Here and in the rest of the paper we use reduced units defined by
the Langevin dynamics and DNA model. The unit of energy is $\epsilon=k_{B}T$,
where we set Boltzmann's constant $k_{B}$ to unity. The bead-to-bead distance
along a single strand defines the unit of length $\sigma$. The mass is $m=1$
for all beads. A Langevin unit of time is defined as $\tau_{L}=\sigma\sqrt{m/\epsilon}$.

\begin{figure}
\centering
\includegraphics[width=0.8\textwidth]{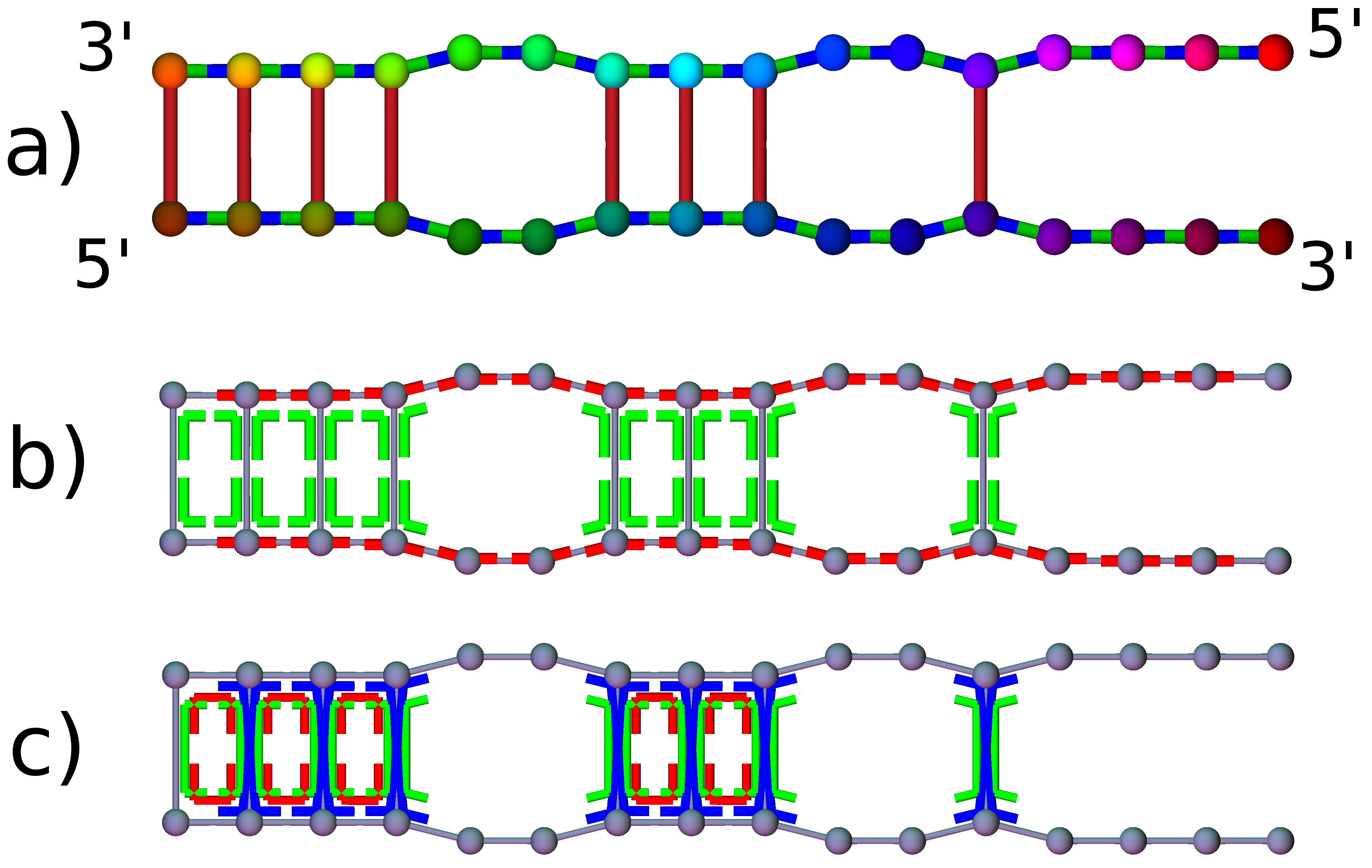}

\caption{Illustrative DNA conformation with partial hybridization. a) nucleotide beads
(different hues indicate complementarity), backbone bonds (green-blue indicates 3'-5' bond
direction), and hybridization bonds (red); b) angular interactions indicated by lines
parallel to the two bonds involved (red/green colours indicate the type of interaction);
c) dihedral interactions indicated by lines parallel to the three involved bonds
(red/green/blue colours indicate the type of interaction). The figure is explained in
the text.}
\protect\label{fig:DNA-model}
\end{figure}

Fig.~\ref{fig:DNA-model}a shows complementary nucleotide beads with
the same hue but different levels of colour saturation. As a simplification,
we allow each bead only to hybridize with a single complementary bead.
The DNA model has two types of bond interactions: permanent backbone
bonds (shown green/blue) and dynamic hybridization bonds (shown red).
Backbone bonds (subscript bb) and hybridization bonds (subscript hyb)
are characterized by the two potentials:

\[
U_\text{bond,bb}(r)=\frac{U_\text{min,bb}}{(r_\text{c}^\text{b}-r_{0}^\text{b})^{2}}\left((r-r_{0}^\text{b})^{2}-(r_\text{c}^\text{b}-r_{0}^\text{b})^{2}\right),
\]
and
\[
U_\text{bond,hyb}(r)=\begin{cases}
\frac{U_\text{min,hyb}}{(r_\text{c}^\text{h}-r_{0}^\text{h})^{2}}\left((r-r_{0}^\text{h})^{2}-(r_\text{c}^\text{h}-r_{0}^\text{h})^{-2}\right) & \mbox{for}\quad r<r_{c}^\text{h}\\
0 & \mbox{for}\quad r\geq r_\text{c}^\text{h}.
\end{cases}
\]

In the simulations, we use 
$U_\text{min,bb}=100\epsilon$, $r_{0}^\text{b}\equiv1\sigma$,
and $r_\text{c}^\text{b}=1.2\sigma$, $r_{0}^\text{h}=2\sigma$ and $r_\text{c}^\text{h}=2.2\sigma$.
Note that $U_\text{bond,hyb}(r)\leq0$ for all distances. When two non-hybridized beads of
complementary type are within a reaction distance $r_\text{c}^\text{h}$ a hybridization bond is
introduced between them. If they move further apart than $r_\text{c}^\text{h}$ again, the
hybridization bond is broken. The pair-interaction between all beads is given by a soft
repulsive potential, while we use the same potential function for angular and dihedral
interactions. They are given by
\[
U_\text{pair}(r)=A\left[1+\cos\left(\frac{\pi r}{r_\text{c}^\text{p}}\right)\right]\quad\mbox{for}\quad r<r_\text{c}^\text{p},
\]
where we use $A=1\epsilon$ and $r_\text{c}^\text{p}=1\sigma$ in the simulations, and
\[
U(\Theta;\Theta_{0},U_\text{min})=-\frac{U_\text{min}}{2}\left(\cos[\Theta-\Theta_{0}]+1\right),
\]

Along the backbone of single strands we use a permanent angular interaction defined by
$U(\Theta;\Theta_{0}=\pi,U_\text{min}=25\epsilon)$. This determines the persistence length
of single strands. In Fig.~\ref{fig:DNA-model}b backbone angular interactions are shown
as thick red lines around the central bead defining the angle.

In real DNA molecules, the hydrogen bonds between Watson-Crick complementary
nucleotides act together with stacking interactions and the phosphodiester
backbone bonds to give rise to a helical equilibrium structure of
the double strand. In our coarse-grained model, we utilize angular and dihedral
interactions to determine the ladder-like equilibrium structure of our DNA model
and to control the collective zippering dynamics. To control the stiffness of the
double strands  and to ensure anti-parallel 3'-5' alignment of the two single
strands, we have assigned directionality to the backbone bonds~\citep{SvaneborgCPC2012}.
This is also necessitated by the fact that the 3' and 5' carbons of the nucleotide
sugar ring have been merged into one single nucleotide bead. Fig.~\ref{fig:DNA-model}a
shows the backbone bonds coloured green/blue to indicate the 3' and 5' ends,
respectively.

When a hybridization bond is introduced, we also dynamically add angular
interactions between the hybridization bond and the neighbouring backbone
bonds. These angular interactions are characterized by the potential
$U(\Theta;\Theta_{0}=\pi/2,U_\text{min,a})$, which favours a right angle
conformation. When a hybridization bond is broken, concomitantly all the associated
angular interactions are removed. In Fig.~\ref{fig:DNA-model}b the angular
interactions are shown as green lines indicating the angle.

Besides introducing angular interactions, we also dynamically introduce
dihedral interactions. A dihedral interaction involves four beads
connected by three bonds, which defines a particular bond pattern,
where the bonds can either be a hybridization bond, a $3'-5'$ backbone
bond, or a $5'-3'$ backbone bond. Three bond patterns are possible.
The bond pattern corresponding to red dihedrals in Fig.~\ref{fig:DNA-model}c,
is characterized by $U(\Theta;\Theta_{0}=0,U_\text{min,d})$ which favours a
planar (cis) conformation. The bond pattern corresponding to blue dihedrals
is characterized by $U(\Theta;\Theta_{0}=\pi,U_\text{min,d},a=0)$ which favours parallel backbone
(trans) conformation. The last dihedral pattern corresponding to green
dihedrals is characterized by $U(\Theta;\Theta_{0}=0,U_\text{min,d})$ which favours
a parallel (cis) conformation. Note that without the directional backbone
bonds, we would not be able to distinguish between these two latter
dihedral patterns.

During a simulation, at each time we introduce a hybridization bond,
we also introduce up to four angular interactions and up to eight
dihedral interactions -- less if the hybridization bond is at the end of
a strand. Let $\Delta$ be the total decrease in binding energy when 
two beads hybridize inside a chain. We assign one third of this
energy to bond, angular, and dihedral interactions, respectively.
 Hence $U_\text{min,hyb}=\Delta/3$, $U_\text{min,a}=\Delta/12$, and
$U_\text{min,d}=\Delta/18$. This choice does not affect the static
properties of the model, which are determined by the total energy
associated with a conformation, however it does influence the dynamic
properties. We define $\Delta=10\epsilon$ as a reference binding
energy. Since only the ratio $\Delta/T$ enters the partition function
of the model, this effectively fixes the absolute melting temperature
of the stands. From a separate set of simulations at varying strand length
$n$ and temperature, we have determined the absolute strand melting
temperatures as $T_m(n=3)=0.74\epsilon$, $T_m(n=5)=0.99\epsilon$,
$T_m(n=10)=1.29\epsilon$, and $T_m(n=15)=1.49\epsilon$, which assuming
room temperature as a reference energy $\epsilon$ corresponds to a
melting temperature interval of $-72$ to $145$ degrees for this range
of strand lengths.

Fig. \ref{fig:and-model} shows the AND gate design that we are simulating. It
consists of two toeholds of identical length $t$ and two adjacent domains of
length $l$. Note that we do not presently worry about garbage collecting the
blocking strand after it has been displaced \citep{Car:2011}. We have performed
simulations of $(t,l)=(3,10), (3,12),$ $(5,10), (7,8),$ $(7,10),$ and $(7,14)$.
Hence we can make comparisons varying toehold length for fixed total strand length
($n=15$), varying the length of the adjacent domain for fixed toehold ($t=7$), and
varying the length of the toehold for fixed length of the adjacent domains ($l=10$),
respectively. To simulate the AND gate, we study a single strand of all the species
in a simulation box of size $L=40\sigma$. We insert the strands as straight
conformations with a random position and orientation. Initially the blocking strand
and the output signal strand are hybridized to the template molecule. We repeat each
simulation at least ten times and in most cases twenty times starting from
statistically independent initial states to obtain statistics on the observed
hybridization kinetics. To estimate gate fidelity, we have also performed simulations
where only one or none of the input signal strands are present. During a simulation,
information is stored when a bond is created or removed between pairs of beads.
Hence we know the entire hybridization bonding dynamics with $\Delta t$ time resolution.

We can relate the simulation units to experimental units as follows.
Identifying the bead-to-bead distance  with the DNA rise
distance $\sigma=0.33\mbox{nm}$ maps to a strand concentration of $\approx 1 \mbox{mM}$.
This is quite high, but allows us to observe more strand collisions and hence to
better characterize displacement kinetics. At lower concentrations most of the simulation
time would be spent on diffusion of isolated strands, which is not of interest. 
The diffusion coefficient of a DNA model strand is $D(n)=k_B T \Gamma/(m n)$
where $n$ is the total number of beads in a molecule. All simulations are run with
$T=1\epsilon$. This can be equated with the DNA diffusion coefficient of a particular
experimental conditions to obtain a time mapping. Extrapolating the data in Ref.
\citep{TinlandMM1997} yields $\tau_L \approx 1.6 \times 10^{-12}s$ for $n=20$. For
the simulation of the AND gate we integrate the dynamics for $10^8$ steps. This
corresponds to approximately a microsecond of real time dynamics and such a simulation
runs in approximately half a day on a powerful PC. Hence the run time per particle
per step is approximately $1\times10^{-5}s$.

\section{Results\label{sec:Results} and discussion}

\begin{figure}
\centering
\includegraphics[width=0.32\columnwidth]{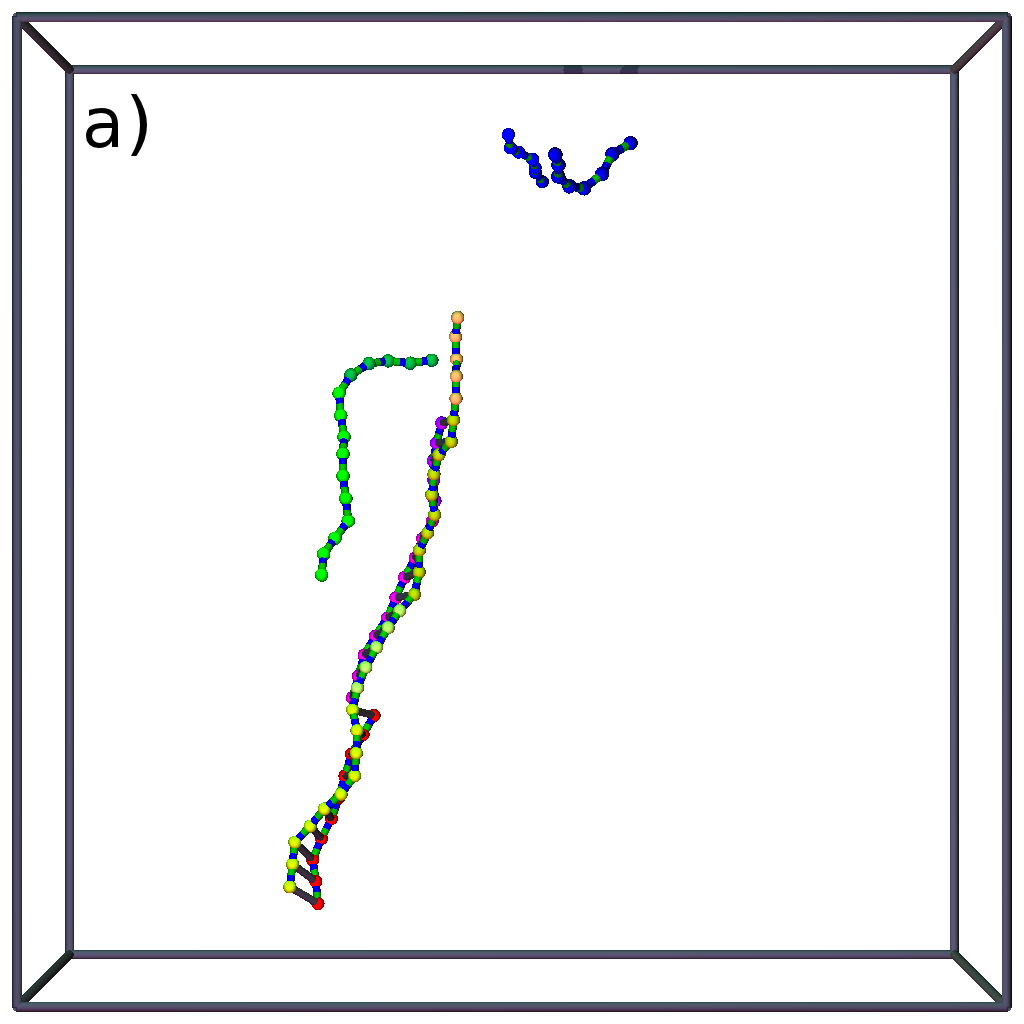}
\includegraphics[width=0.32\columnwidth]{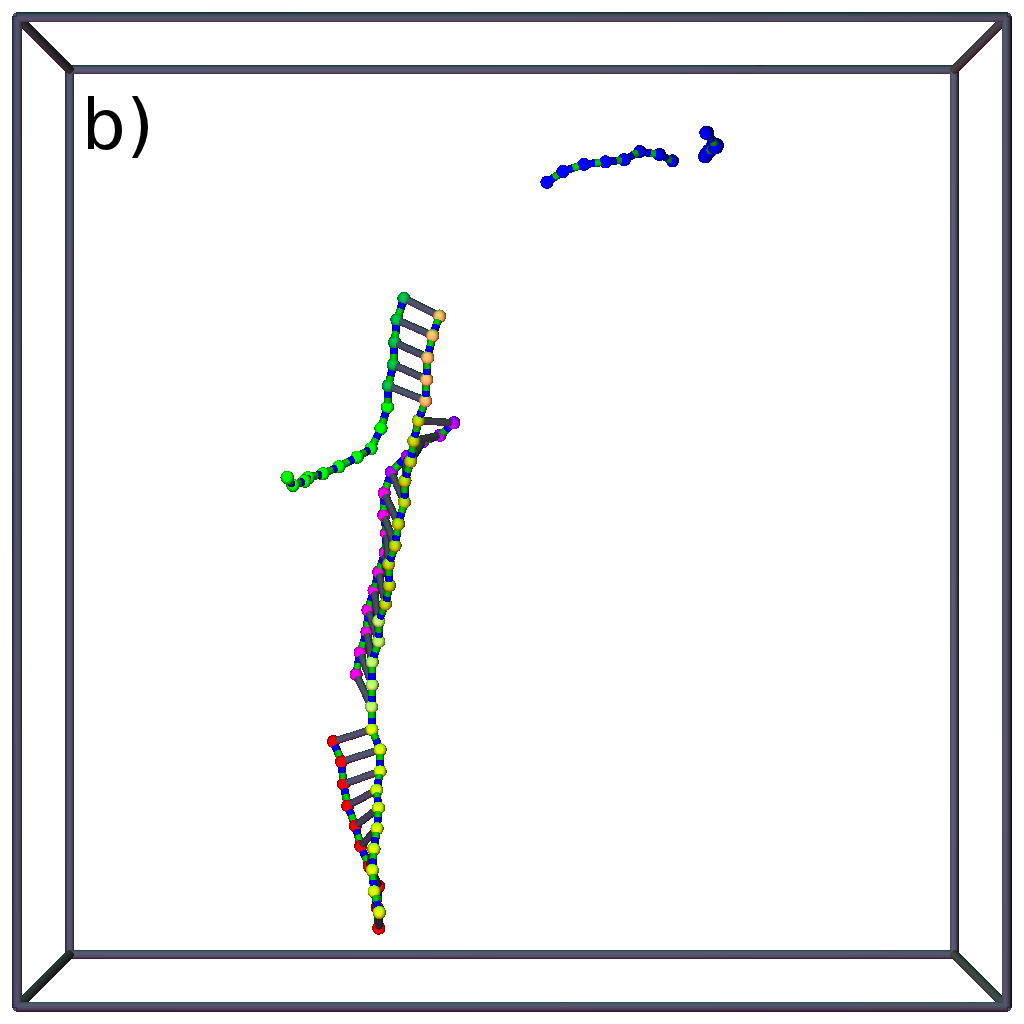}
\includegraphics[width=0.32\columnwidth]{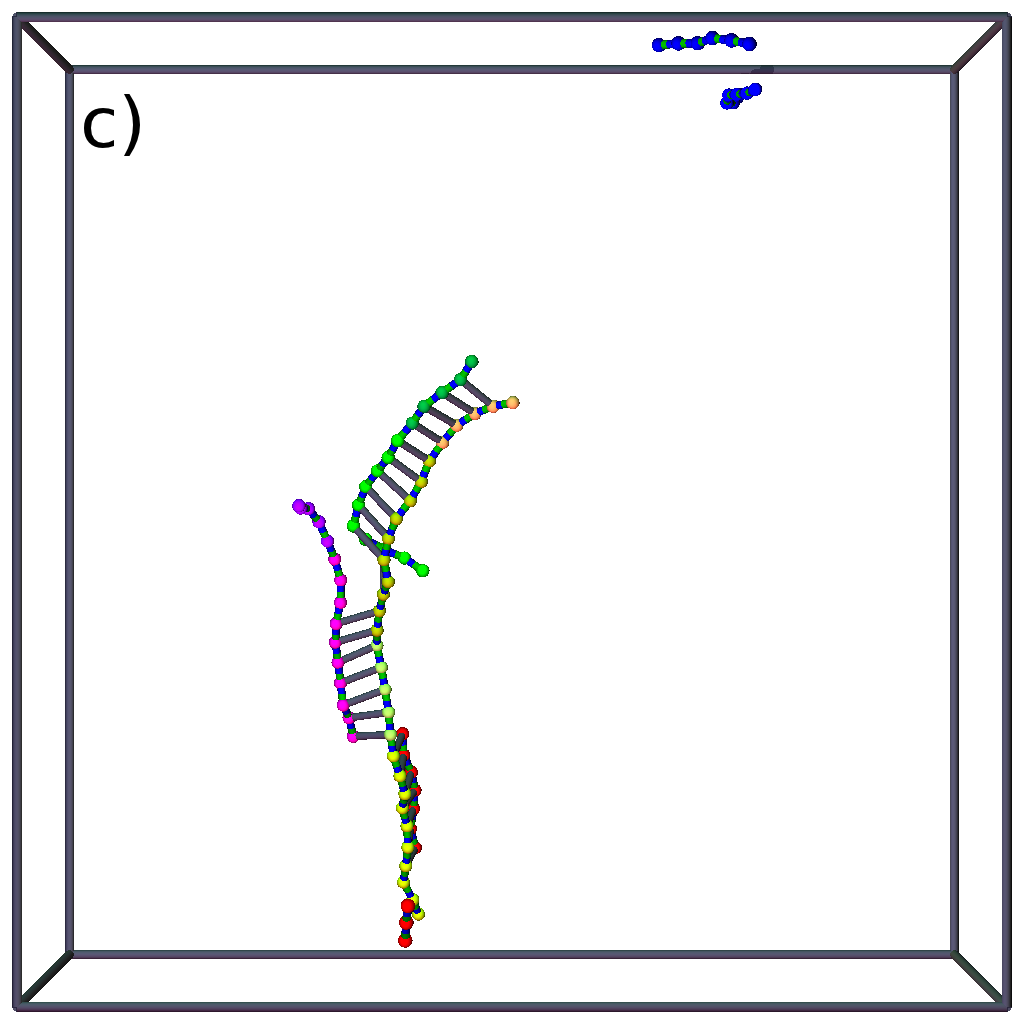}

\includegraphics[width=0.32\columnwidth]{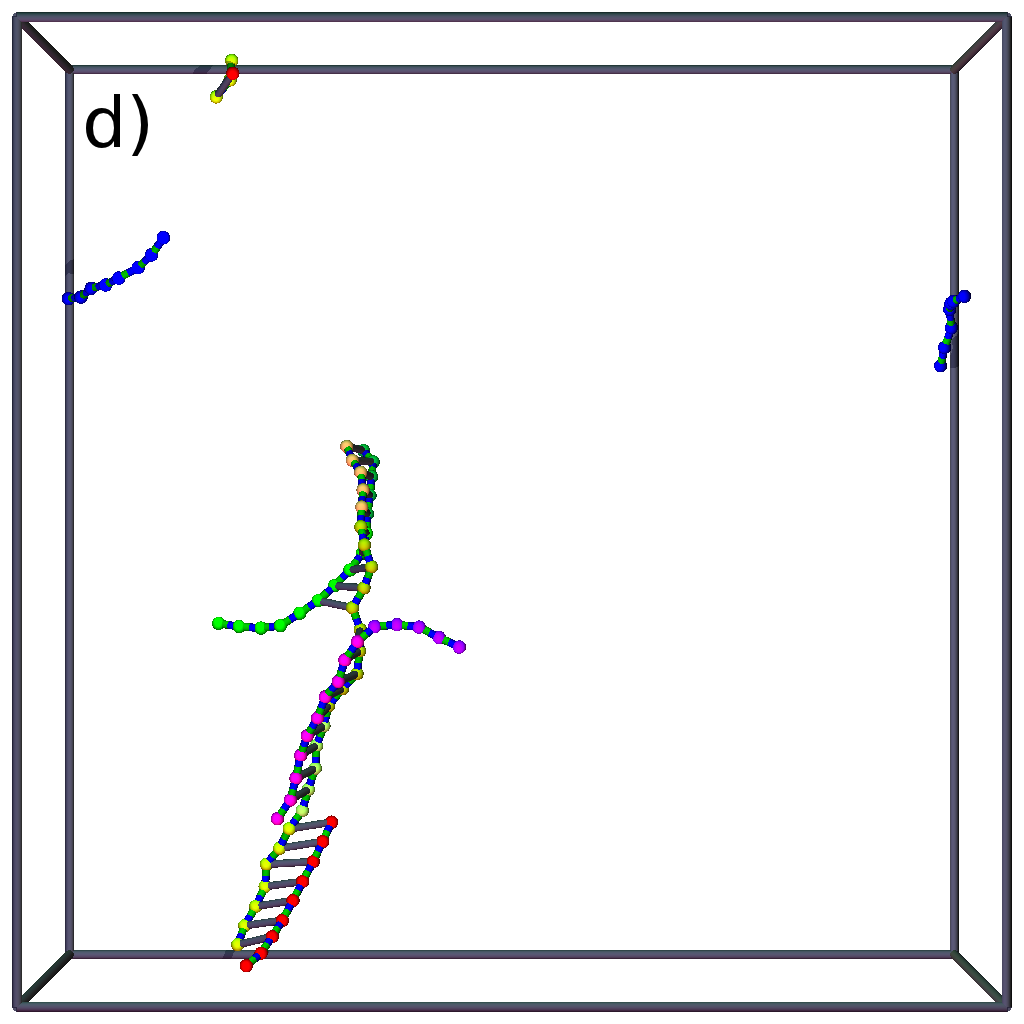}
\includegraphics[width=0.32\columnwidth]{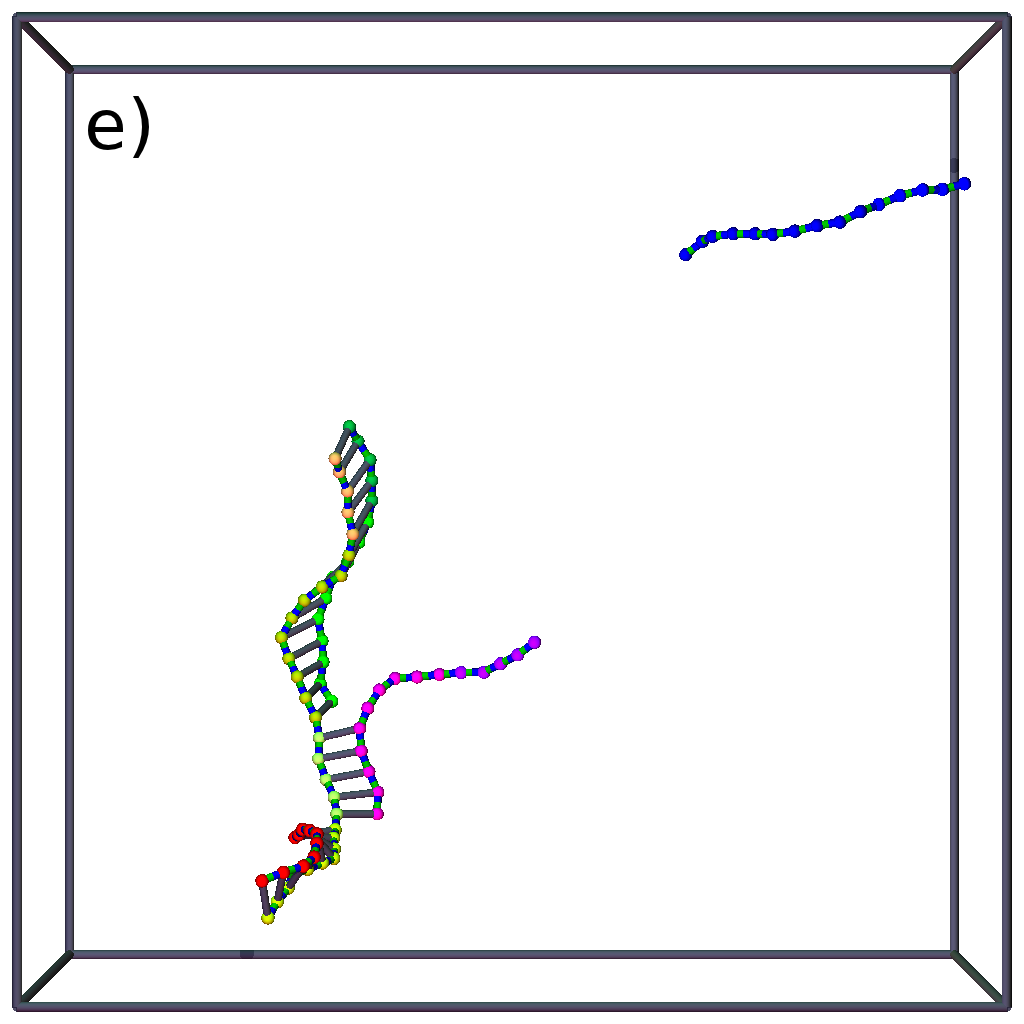}
\includegraphics[width=0.32\columnwidth]{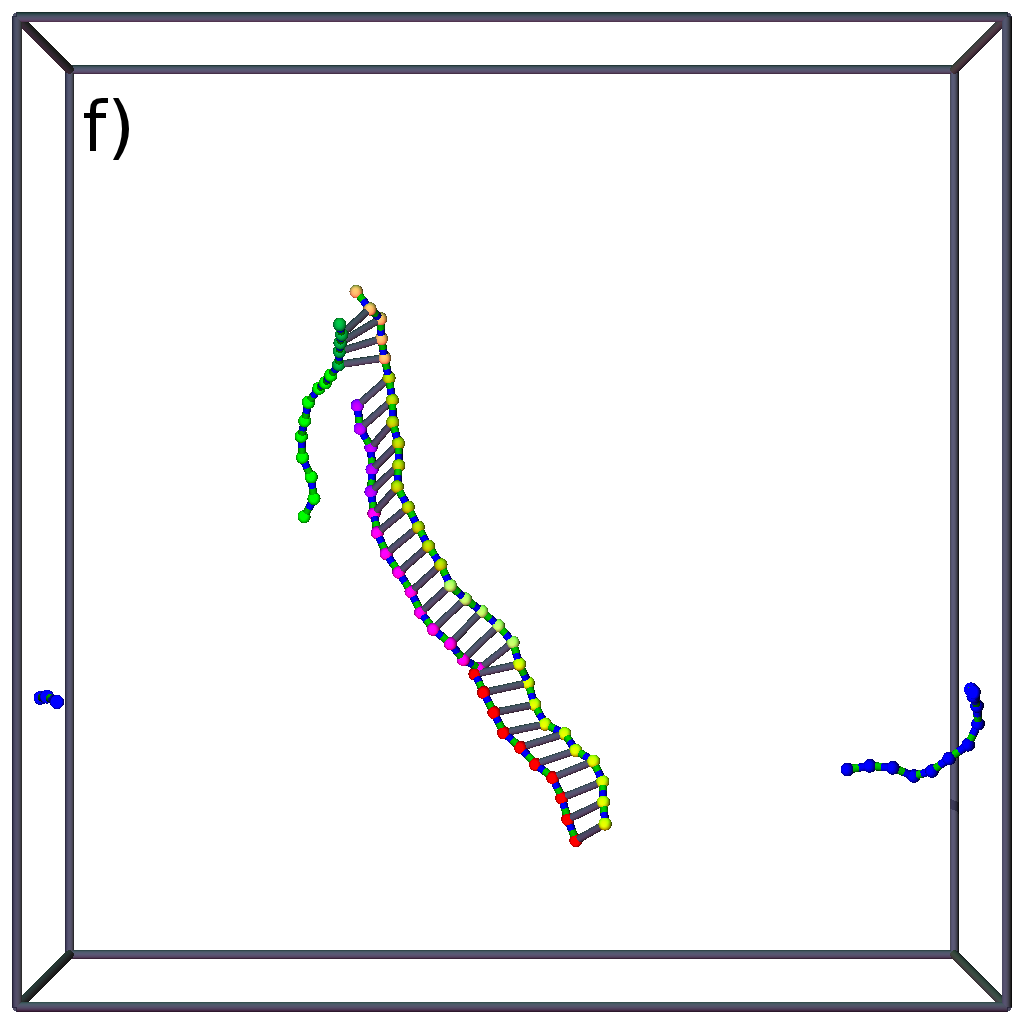}

\includegraphics[width=0.32\columnwidth]{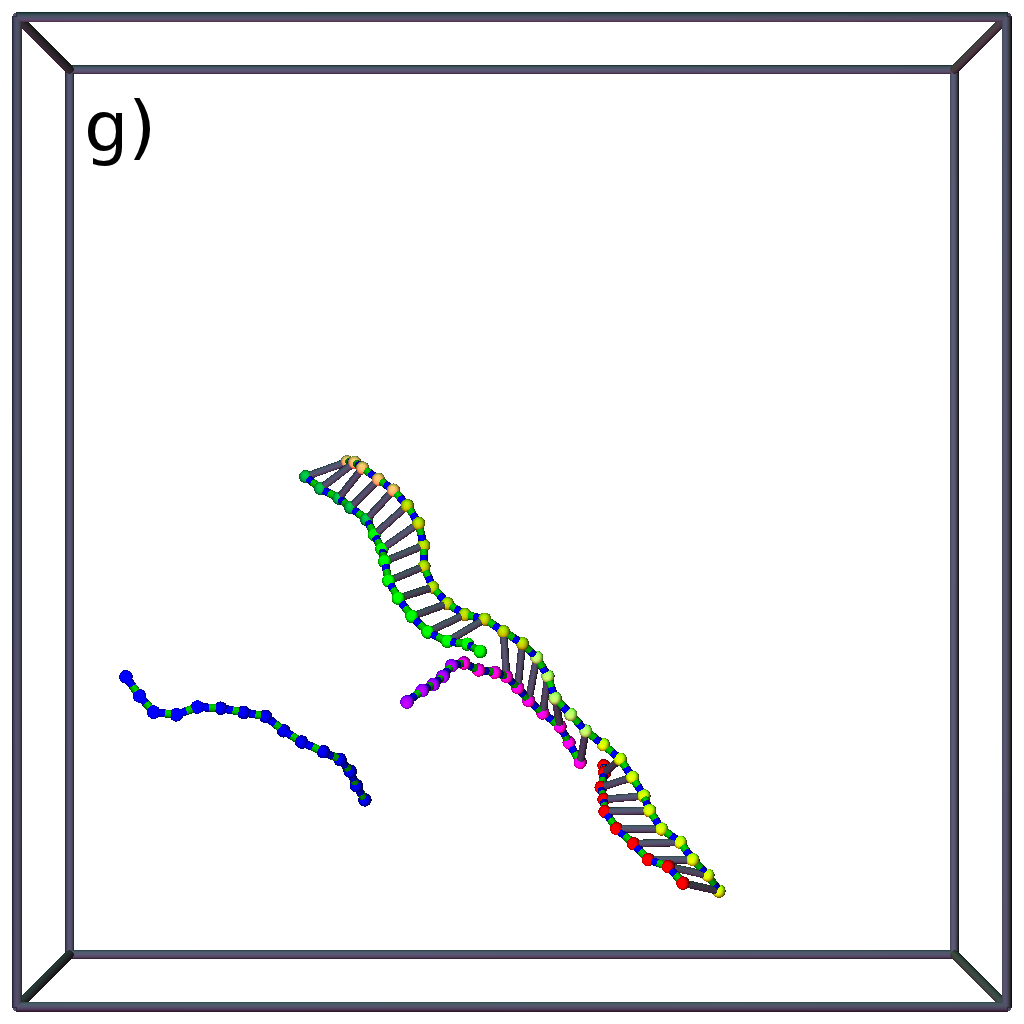}
\includegraphics[width=0.32\columnwidth]{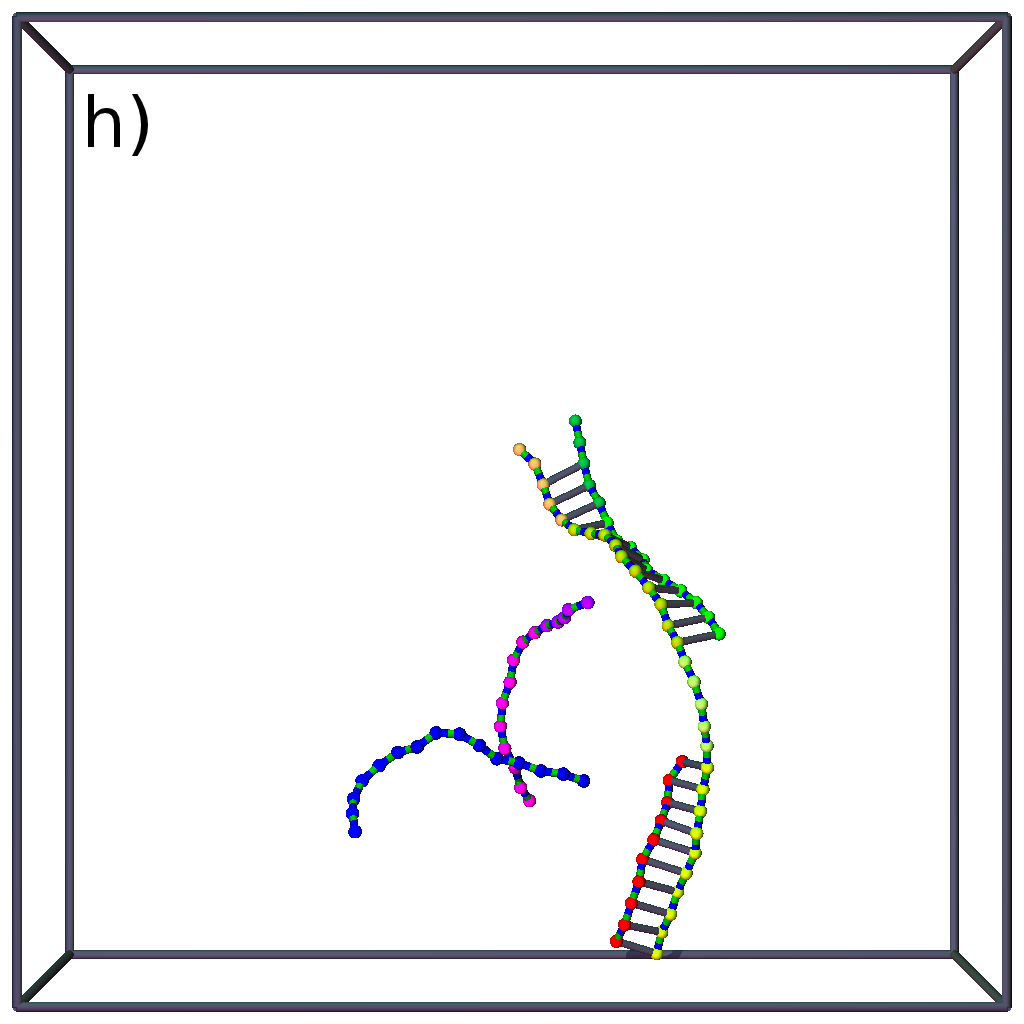}
\includegraphics[width=0.32\columnwidth]{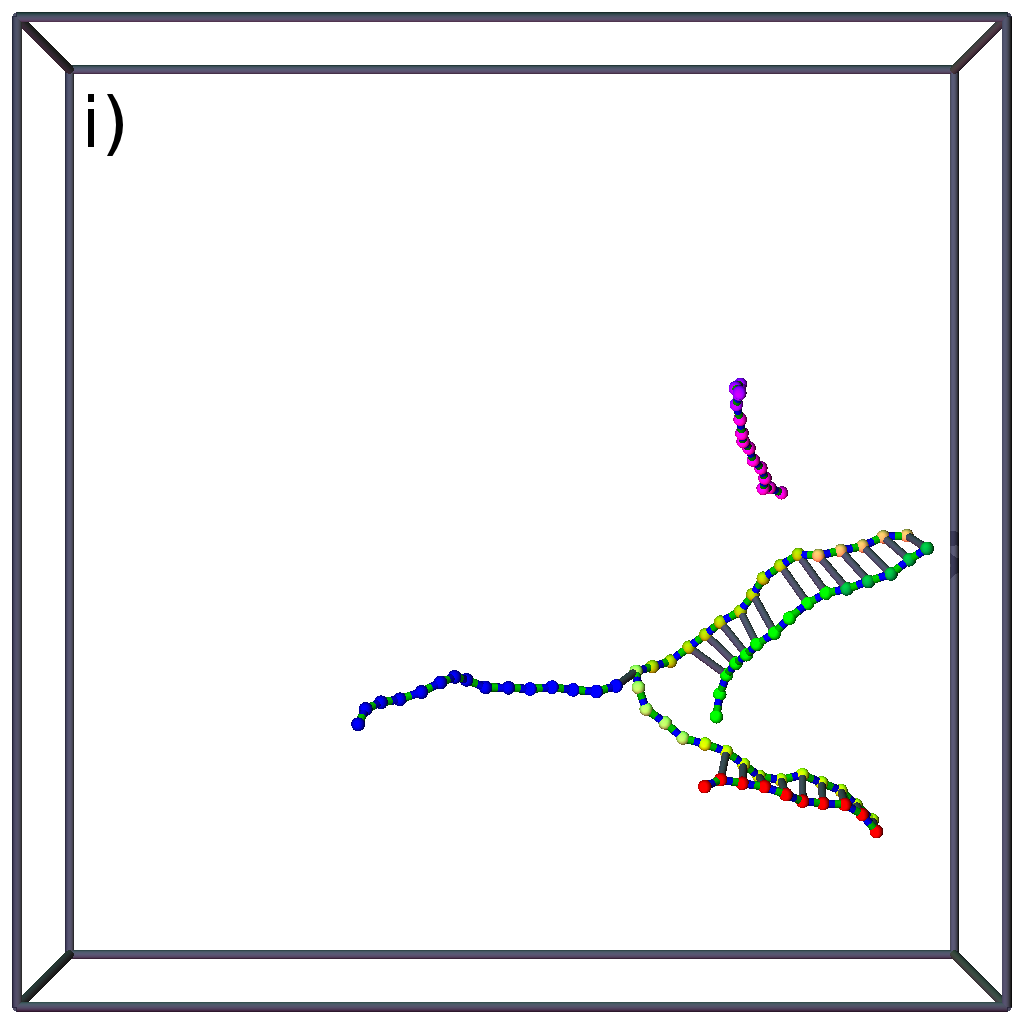}

\includegraphics[width=0.32\columnwidth]{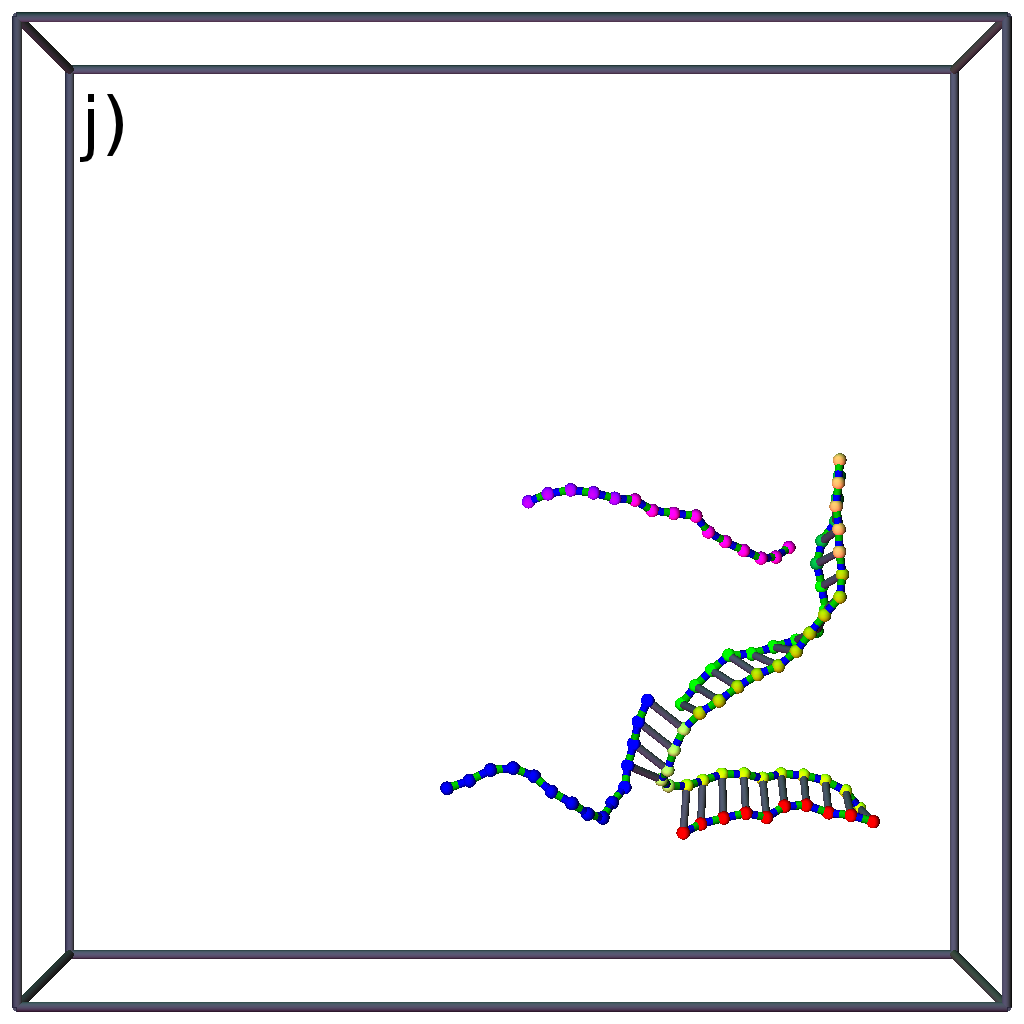}
\includegraphics[width=0.32\columnwidth]{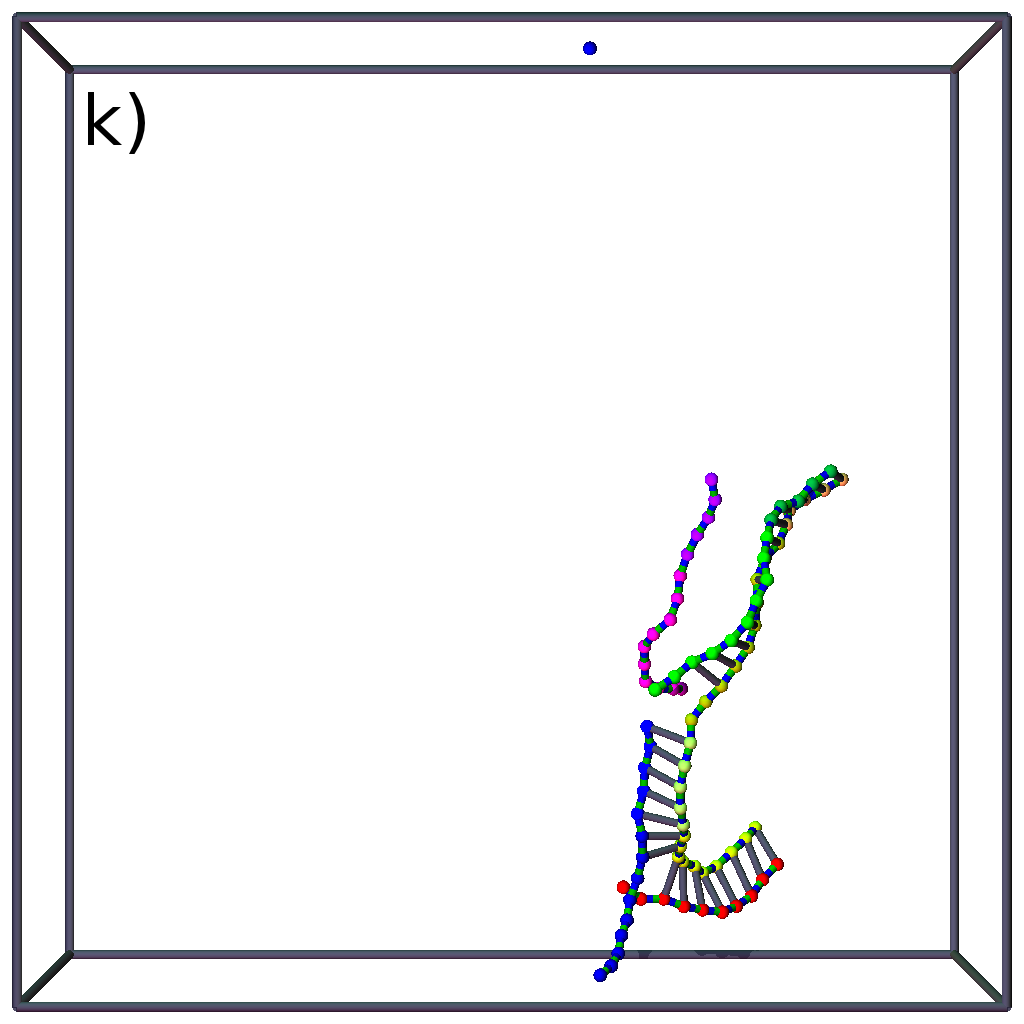}
\includegraphics[width=0.32\columnwidth]{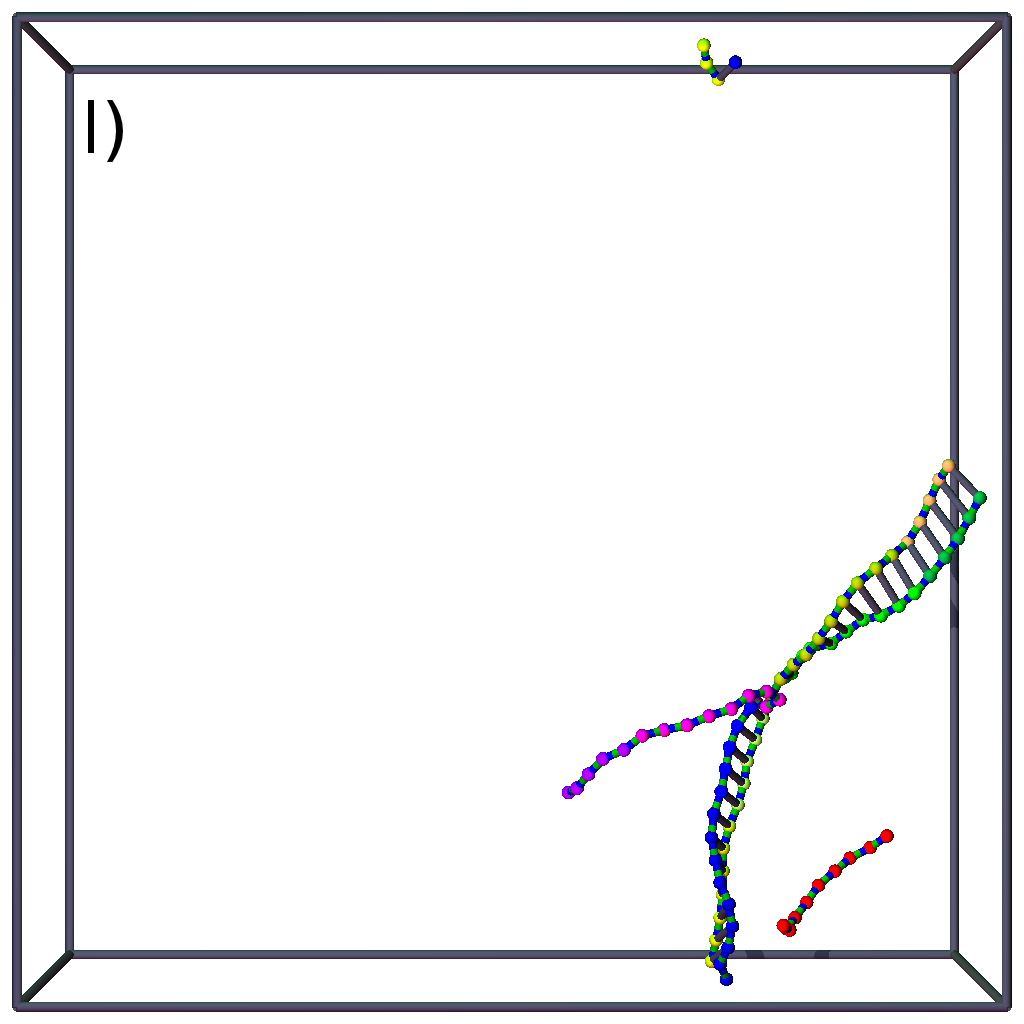}

\caption{\protect\label{fig:Renderings}Snapshots of strand displacement dynamics from a simulation with $(t,l)=(5,10)$.
The colours of the strands match Fig. \ref{fig:and-model}. 
a) signal strand 1 (green) is close to toehold 1. b) hybridization of 
signal strand 1, c-g) diffusion of the branch point between the
signal strand 1 and the blocking strand (magenta),
h) the  blocking strand finally displaced and the second toehold exposed,
i) signal strand 2 (blue) is close to toehold 2, j) hybridization of 
signal strand 2, k-l) displacement and release of the output strand (red).
The individual snapshots (a-l) are taken at times $8305$, $8330$,
$8355$, $8525$, $8675$, $9090$, $9280$, $9280$, $12985$, $13010$,
$13040$, $13170\tau$. Note that we use periodic boundary conditions, hence 
strands leaving one side of the box enter the box on the opposite side.
The simulations these snapshots are
obtained from correspond to the simulation shown in row 7 in the top graph
in Fig. \ref{fig:Time-evolution}.
}
\end{figure}

\begin{figure}
\centering
  \includegraphics[width=0.7\columnwidth,angle=270]{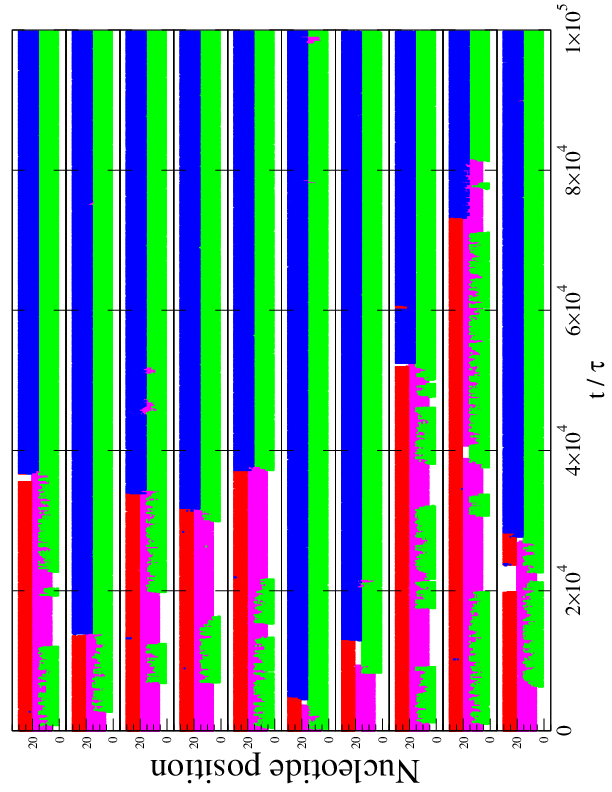}
  \includegraphics[width=0.7\columnwidth,angle=270]{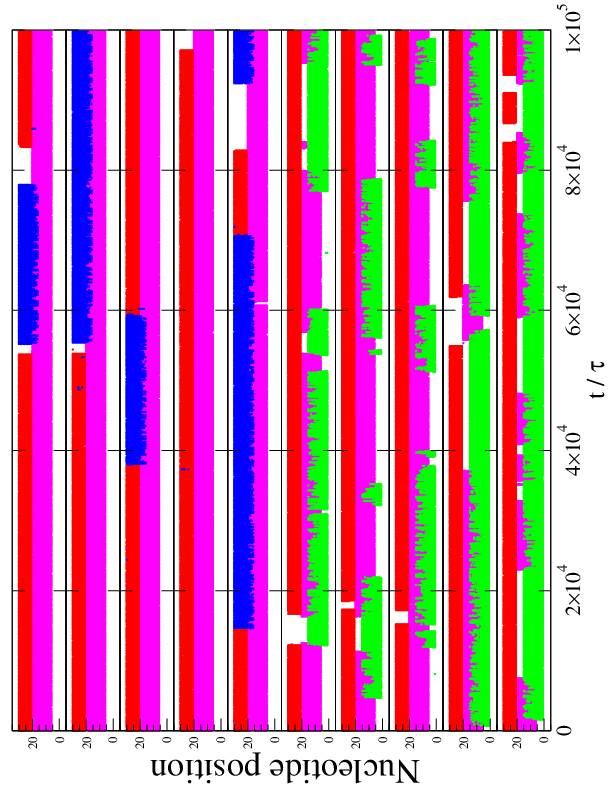}

\caption{\protect\label{fig:Time-evolution}
Strand displacement dynamics for statistically independent simulations
with $(t,l)=(5,10)$ where both the signal strands (green and blue)
are present (ten rows in the upper graph), and where only the second
or the first signal strand (blue or green) is present (top five and bottom
five rows in the lower graph, respectively) The hybridization state of each
nucleotide along the template is shown colour coded along the y axis:
blocking strand (magenta), first signal strand (green), second signal
strand (blue), and the output strand (red).
}
\end{figure}

Fig.~\ref{fig:Renderings} shows simulations of the strand displacement
process of the AND gate (c.f. Fig.~\ref{fig:and-model}) using our Dynamic Bonded DNA model. As desired, the
gate releases the AND signal (red) after binding both input strands (blue
and green). Fig.~\ref{fig:Renderings} illustrates the dynamic nature of
the strand displacement process as required by a single AND computation.
Throughout the paper we will refer to the strands by the colour they have
in Fig. \ref{fig:and-model}. This colour coding is also used in the plots
when referring to single strands. Fig.~\ref{fig:Renderings}a-h shows how the
green signal strand hybridizes with the template, and how the branch point
between the green signal strand and the magenta blocking strands diffuses
forwards and backwards until it finally displaces the magenta blocking strand. In the final
steps (i-l), the blue signal strand hybridizes with toehold 2 and displaces
the red output stand. This completes the strand displacement transitions
constituting the AND computation.
From the simulations in Fig.~\ref{fig:Renderings}, we can also measure
the durations of the transitions involved in the gate computation. The
total transition time (a-l) of the AND gate is $4865\tau$ from the initial
binding of the green signal strand, subsequent binding of the blue signal
strand until the final displacement of the red output strand. Whereas the
first displacement transition (a-h) takes $975\tau$, the final displacement
transition (i-l) takes just $185\tau$. In order to understand this kinetics,
we must understand the physical dynamics of the AND gate in more detail.

In the first displacement transition (a-h), the initial (a) and final (h) 
states have essentially the same energy since the two toeholds have the same
length. As a result the displacement process is completely reversible.
We observe that the branching point between the two strands can diffuse
backwards and forwards since when one of the strands is completely hybridized
with the template, the other strand is still hybridized with a toehold.
This competition for the domain first stops when one of the two strands is
released from its toehold due to thermal melting. It is worth noting that
a strand that is thermally irreversibly bound to a template can nevertheless be
displaced by another strand with a longer complementary sequence. 
In the final displacement process (i-l), we observe the blue signal strand
displacing the red output strand. This process is asymmetric since the
red strand is only hybridized to the domain adjacent to toehold 2 and
does not have a third toehold to hold on to. Hence when the blue signal strand
has completely hybridized to the template, the red strand is released. This
is the state with the minimum energy, since the system has decreased its
total energy by the binding energy of toehold 1.

Fig.~\ref{fig:Time-evolution} shows $20$ instances of the strand displacement
dynamics of the operation of the AND gate. Note that the hybridization state
is only shown for every $100\tau=10^5\Delta t$, hence the plot only shows
the most long-lived transitions. The upper graph shows the operation
of the gate when both signal strands are present, whereas the bottom graph 
shows the gate operation when only the blue or the green signal strands are
present (corresponding to the 11, 01, and 10 logic states of the gate
respectively). We observe that the red output signal is irreversibly released
in all simulations where both input strands are present. The computation is
energetically downhill and driven by the binding energy of the first toehold
with the green input strand. 
The lower graph in Fig.~\ref{fig:Time-evolution} shows the operation of the
gate when only one of the two input strands is present. In the case where only the green
strand is present, we see that it readily hybridizes with the template and
competes with the magenta strand. In the case where only the blue strand is
present, we see transitions where the blue strand hybridizes with the template
and erroneously displaces the red signal strand.
Both in simulations with one or both signal
strands present, we observe transitions where the red output strand is
erroneously released due to thermal melting of domain. When the blue
strand is present we see that it displaces the red strand even though the
second toehold is blocked by the magenta strand. 
We also observe that the red signal strand is eventually released in all
simulations where both input strands are present. The red signal strand
remains hybridized in most of the simulations where only the green input
strand is present. However, we observe that the red signal strand is released quite
often when then the blue input strand is present. Below we will quanitify
the gate fidelity for all the simulations and discuss this further.

We can also confirm that the dynamics we observed in the snapshots in
Fig.~\ref{fig:Renderings} is indeed representative for the general strand displacement
dynamics of the AND gate. Fig.~\ref{fig:Time-evolution} shows that there is
strong competition between the green signal strand and the magenta blocking
strand for hybridization with the domain 1, and the branching point
can diffuse on the domain for an extended period of time. The diffusion
process stops when one of the two strands leaves its toehold due to thermal
melting. Hence the duration of these processes should depend exponentially
on the length of the toehold. We also observe that the displacement of the
red strand by the blue signal strand is a much faster process, which can
hardly be resolved in the figure.

\begin{figure}
\centering
  \includegraphics[angle=270,width=0.80\columnwidth]{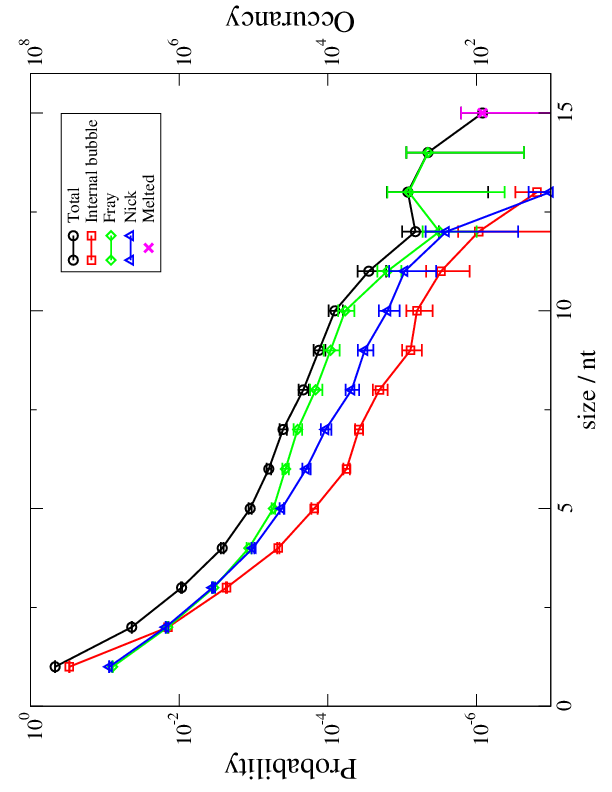}

\caption{\protect\label{fig:defect}
Size probability distribution of all defect types (black circles), and the contributions
from internal bubbles (red boxes), frays (green diamonds), nicks (blue triangles) for a
template of length ($n=30$) and two strands of length ($n=15$) in a simulation of temperature $T=1\epsilon$.
}
\end{figure}

Fig.~\ref{fig:Time-evolution} shows
that the simple ``textbook'' pathway illustrated in Fig.~\ref{fig:and-model} is
by no means the only or even the most likely pathway from the initial to the final
state of the AND gate. Although the final state is reached with good fidelity,
there are many possible pathways that the system can follow to reach this state.
These simulations illustrate the less than ideal behaviour of strand displacement
gates that will occur when they are implemented in real chemistry.
At non-zero temperature a DNA strand will always have transient stretches or bubbles
of thermally broken hybridization bonds as these contribute
configurational entropy, and hence reduce the free energy.\citep{PhysRevE.68.061911}
In the case where a such a bubble is at the free end of the template
it is denoted a fray, whereas we denote it a nick when the bubble originates
at the interface between two strands hybridized to the template. These defects can transiently
expose long stretches of nucleotides where a free complementary strand can bind
and transiently or permanently displace the original strand.

To characterize the occurrence of bubbles, frays, and nicks in our DNA AND gate,
we have performed simulations of the final state corresponding to Fig. \ref{fig:and-model}f
without the displaced magenta and red strands being present in solution. These
simulations have been run up to $t=10^5\tau$ as the other simulations. At each time
step, we analyse the blue and green strand conformations for internal bubbles, frays
at the ends of the template, or nicks at the interface between the green and blue
strands. The corresponding probability distributions are shown in \ref{fig:defect}.
We observe an approximately exponential decrease
in defect probability as the defect length increases. In 50\% of the sampled
conformations a single hybridization bond is found to be broken, which is most likely
an internal bubble. Already defects of three nucleotides and longer are significantly
more likely to be frays or nicks, and the largest defects are predominantly frays.
This is expected since nicks and in particular frays have less stabilizing
angular and dihedral interactions compared to a nucleotide inside a double strand.
While the probability of these large defects is small, the simulation is long enough
that their occurrence is significant. If such a defect occurs while a longer
complementary strand is within hybridization distance of the template, then
these thermally induced defects will cause undesired strand displacement transitions.

\begin{figure}
\centering
  \includegraphics[width=0.80\columnwidth]{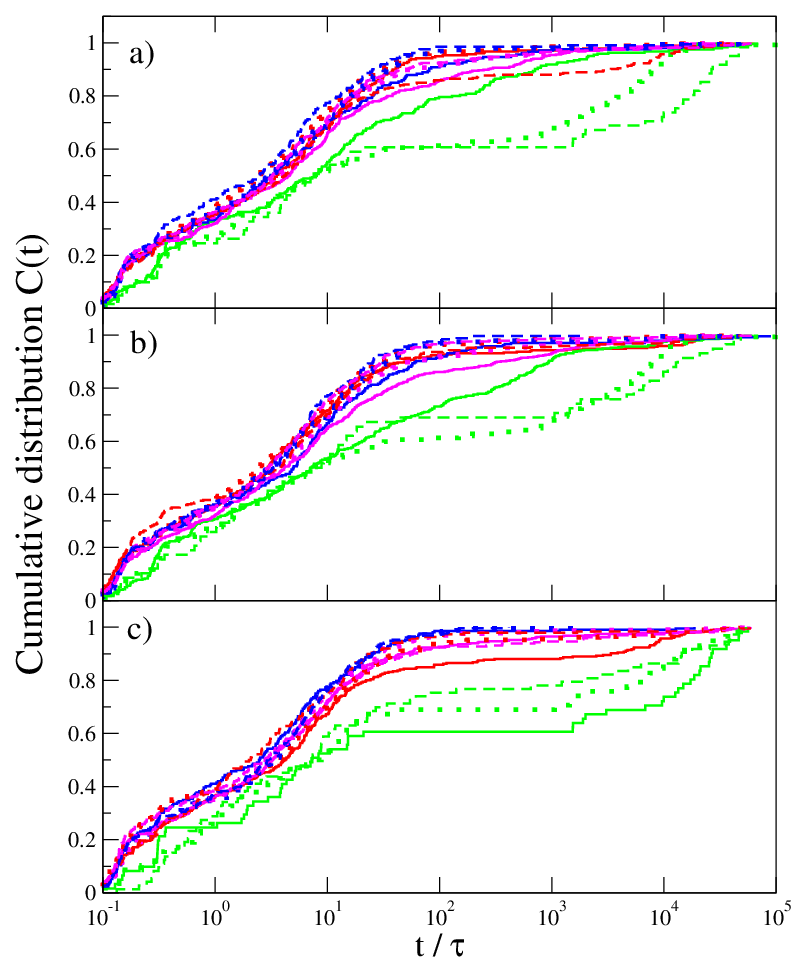}

\caption{\protect\label{fig:Onoffdynamics}
Cumulative probability distributions for the duration of hybridization$\rightarrow$release
transitions of each strand for simulations with a) constant strand length
$(t,l)=(3,12), (5,10), (7,8)$, b)
varying toehold size
$(t,l)=(3,10), (5,10), (7,10)$,
and c)
varying domain size
$(t,l)=(7,8), (7,10), (7,12)$ using solid, dotted, and dashed lines, respectively,
for the three systems. The colours denote the strand type: signal strand 1 (green),
signal strand 2 (blue), blocking strand (magenta), and output strand (red).
}
\end{figure}

To analyse the strand displacement dynamics in more detail, we measure
the duration of each hybridization$\rightarrow$release transition of each single strand.
This is the time from when the first hybridization bond between a particular strand
and the template is created until the last hybridization bond is broken and the strand is released.
Note that these do not need to coincide with the first
or last nucleotides of the respective domains. E.g., for the green strand transition, pathways
such as b$\rightarrow$a, b$\rightarrow$c$\rightarrow$b$\rightarrow$a, and
b$\rightarrow$c$\rightarrow$d$\rightarrow$c$\rightarrow$b$\rightarrow$a in
Fig. \ref{fig:and-model} all contribute to the hybridization $\rightarrow$ release
transitions. The large number of possible transition pathways already suggests that we
can expect a very broad time distribution of these transitions. Note that the first
release of the magenta and red strands is not sampled, since they are already hybridized
with the template when the simulation starts. Equally, incomplete transitions
at the end of the simulation are also discarded.

In the following plots, we show the cumulative distributions $C(t)=\int_0^t P(t')\mbox{d}t'$
rather than the probability density $P(t)$. Our measured probability densities
are of the form $P(t)=N^{-1}\sum_i^N \delta(t-t_i)$ where $\delta$ is the Dirac delta function
and $t_i$ is the duration of hybridization $\rightarrow$ release event $i$ we obtained from
simulation. To represent
this distribution as a continuous distribution would require binning,
which could potentially introduce artefacts in the representation. Instead, the cumulative 
distribution is increased by $N^{-1}$ every time an event occurs. In this representation a
high density of events i.e. a peak in a continuous probability distribution produces a
high slope in the cumulative distribution, whereas a flat plateau indicates
zero probability density of events. Below by peaks we denote regions of high slope in the
cumulative distributions. The median transition time $\langle t \rangle$ is trivially
read of at $C(\langle t \rangle)=0.5$.

Ideally, if the gate follows the pathway indicated in Fig.~\ref{fig:and-model},
we would not observe a single hybridization $\rightarrow$ release transition.
The blue and green signal strands would bind irreversibly and never be released.
The magenta blocking strand and the red output strand would be released only
once, but they would never rehybridize with the template. Fig.~\ref{fig:Onoffdynamics}
shows the cumulative distributions of these transitions for each of the four strands
constituting the operation of the AND gate when both signal strands are present.
Clearly, we see a large number of transitions occurring for all the strands involved
with the gate computation.
Furthermore, we observe an extremely wide distribution of transition times covering
at least six orders of magnitude in time. Since the simulations are run up to $10^5\tau$,
this imposes an upper limit on the observable transitions. It is likely that
a fraction of transitions would occur on even longer time scales.

Overall, the distributions shown in the figure are very similar. Most of the
fast transitions of the blue signal strand and the magenta blocking strand occur
on time scales of $10^{-1} \cdots 10^2\tau$, although about $10\%$ of the transitions
are in the long tail of the distribution. We also observe a large number of
slow transitions for the green signal strand that occurs on time scales of
$10^{-1} \cdots 10^5\tau$. For most of the simulations, except for the one with the
shortest domain length, we also observe fast transitions for the red output
strand. Note that only slow transitions with $t \gg 10^2\tau$ are visible in Fig.~\ref{fig:Time-evolution}
For the fast transitions, the distributions are quite similar, and they correspond to
a distribution with two peaks one centered around $t\sim 10^{-1}$ and $10^1\tau$.
The first peak probably corresponds to binding events where one or a few hybridization
bonds are made to nucleotides exposed within defects, but where the invading strand is
rapidly displaced. The second peak probably corresponds to binding events where most of
the strand hybridizes with the template, but is subsequently displaced by another strand.
Alternatively, this can be interpreted in light of the free energy of the transition.
The hybridization binding energy favours the double stranded state, however, the entropy
favours having a pair of single strands,
since this maximise the number of possible conformations of the system. The double
peak is suggestive of an free energy barrier of hybridization, where the reduction
in free energy due to a few initial hybridization bonds have not yet been balanced
by the loss in entropy due to the loss of possible conformations.
For the green strand, a third broad peak appears at
very large time scales, and $20-40\%$ of the transitions are long-lived. These are
the reversible strand displacements where the green signal strand and the magenta
blocking strands are competing for the template, and where both are hybridized to
their respective toeholds. With this in mind, one would naively expect the same
transition time distribution for the magenta blocking strand and the green signal
strand, since their displacement process is energetically reversible. However,
the magenta blocking strand is also competing for toehold 2 with the blue strand.
When the blue strand hybridizes with toehold 2 and displaces the red output strand,
this is the state with the minimal energy. Hence the magenta blocking strand is
prevented from rehybridizing with toehold 2 and displacing the green signal strand.
This explains why the transitions of the magenta blocking strand take about the
same time as the green signal strand but are much rarer.

Looking at the kinetic dependence on the lengths of the toeholds and domains, we
see a strong influence on the transitions of the green strand, but relatively
little influence on the transitions of the other strands as long as the toehold
is short compared to the adjacent domain. We observe a dramatic slowing down of
the transitions of the green signal strand as the length of toehold 1 is
increased, whereas the duration of the transition remains around $10^4\tau$ when the
toehold length is kept fixed. If we want to minimise the duration of the green
strand displacement transitions, we should instead increase the domain length
compared to the toehold length as shown by the two simulations with $(t,l)=(3,12)$ and $(3,10)$.

\begin{figure}
\centering
  \includegraphics[width=0.99\columnwidth]{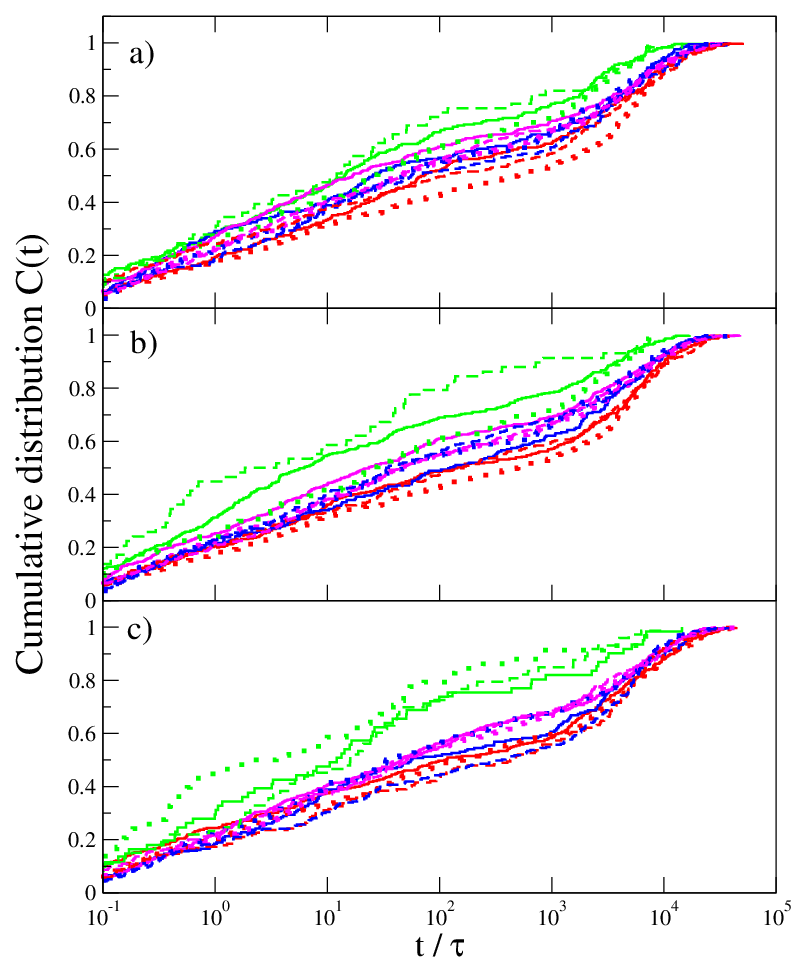}

\caption{\protect\label{fig:Offondynamics}
Cumulative probability distributions for the durations of release $\rightarrow$
hybridization transitions of each strand for simulations with a) constant strand length
$(t,l)=(3,12), (5,10), (7,8)$, b)
varying toehold size
$(t,l)=(3,10), (5,10), (7,10)$,
and c)
varying domain size
$(t,l)=(7,8), (7,10), (7,12)$ using solid, dotted, and dashed lines, respectively,
for the three systems. The colours denote the strand type: signal strand 1 (green),
signal strand 2 (blue), blocking strand (magenta), and output strand (red).
}
\end{figure}

Just as we can measure the durations of all the hybridization $\rightarrow$ release
transitions, we can also measure all the release $\rightarrow$ hybridization transitions.
Whereas the former characterize the cooperative and competitive kinetics of the
hybridization processes, the latter characterize bulk diffusion between
these binding events. As a result we expect these to depend strongly on the
concentration, whereas the hybridization $\rightarrow$ release transitions are not
expected to be strongly dependent on concentration. Fig.~\ref{fig:Offondynamics} shows
the distribution
of the release $\rightarrow$ hybridization transition times. We see relatively little
structure in these cumulative distributions: they are approximately linear over three to
four orders of magnitude with a peak around $10^4\tau$. The approximately linear relation
between cumulative probability and the logarithm of time seen in the figure corresponds
to a transition probability distribution $P(t)\propto t^{-1}$ for $0.1<t<10^3\tau$.
The peak is most likely an artefact due to the periodic boundary conditions that we apply
in the simulations. When a strand is released it will diffuse in free space and return to
the template. However, it can also return to any of the periodic images of the template.
These periodic images define a cubic lattice with spacing $L=40\sigma$. If the free strands
could hybridize with each other or with garbage collecting DNA templates, we would expect
to see more structure in these transitions. Surprisingly the green strand strand exhibits
faster release $\rightarrow$ hybridization transitions than the magenta strand. 
This indicates that the binding kinetics of the first and second toehold are
in fact different, which could be due to the presence of the blue strand, or the fact
that toeholds at the end and internally in a strand are not equally accessible.


\begin{figure}
\centering
  \includegraphics[width=0.99\columnwidth]{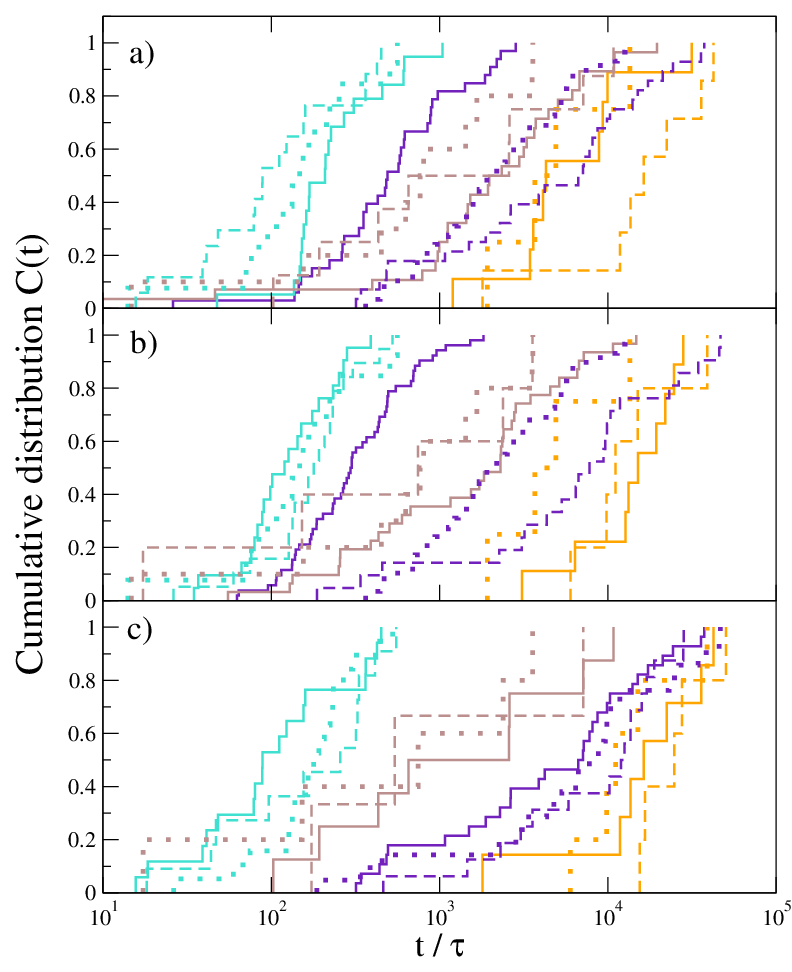}

\caption{\protect\label{fig:displacementdynamics}
Cumulative probability distributions for the duration of the gate operation and
its three constituting displacement transitions for simulations with  a) constant strand length
$(t,l)=(3,12), (5,10), (7,8)$, b)
varying toehold size
$(t,l)=(3,10), (5,10), (7,10)$,
and c)
varying domain size
$(t,l)=(7,8), (7,10), (7,12)$ using solid, dotted, and dashed lines, respectively.
The colours denote the type of transition:
hybridization of the green signal strand $\rightarrow$ release of the magenta blocking strand (indigo),
release of the magenta blocking strand $\rightarrow$ hybridization of the blue signal strand (light brown),
hybridization of the blue signal strand $\rightarrow$ release of the red output strand (turquoise),
and total AND gate computation transition time (orange).
}
\end{figure}

Of key interest is the average time of the total AND computation and its three constituting transitions.
Therefore, we measure the time between the first hybridization of the green signal until the final
release of the magenta blocking strand (Fig.~\ref{fig:and-model} b to c).
Similarly, we measure the time between the release of the magenta blocking strand and the
first hybridization of the blue input strand, as well as the time between the first 
hybridization of the blue strand and the final release of the red signal strand.
In light of the complex sequence of transition events that we observe in
Fig. \ref{fig:Time-evolution}, an non-trivial question is how to define the pattern
of transition events that most accurately characterize the duration of the full
computation transition.
We choose to start (or reset) the clock when the green input strand hybridizes with
the template in the input state  (Fig. \ref {fig:and-model}a), i.e. where the magenta
blocking strand and the red output strand are hybridized with the template. The
first milestone is the release event of the magenta blocking strand. The second milestone
is the hybridization of the blue input signal. We stop the clock the first time the
red output stand is released (e.g. Fig. \ref {fig:and-model}f). Note that this
event pattern specifies a certain causal ordering of milestones, but it does not define a
unique sequence of consecutive events, since we allow hybridization $\rightarrow$
release events between the specified milestones.
Furthermore, in the final state we
do not require the magenta blocking strand to be in solution, nor that the green and
blue input strands remain hybridized with the template. This definition is a compromise
between an ideal, maximally restrictive definition and a more pragmatic definition.
The latter definition allows to analyse our realistic simulation data in a meaningful
way. It is worth noting that we have not observed a single AND computation
characterized by the maximally restrictive definition.

Fig.~\ref{fig:displacementdynamics} shows the
cumulative distributions of the two strand displacement transition times, the
binding transition time and the total gate transition time. The fastest strand
displacement process is the final displacement of the red output strand by the
blue signal strand.
The third transition takes $\approx 10^2\tau$ and is energetically downhill when
the second toehold is exposed. We observe that increasing the domain length
leads to somewhat slower displacement times, while varying the toehold length
has little influence on the displacement speed.
This is consistent with the increased time it takes the branching
point to diffuse across a longer domain after the blue input strand has hybridised
with the second toehold.
The second transition is the binding of the blue input strand after the
magenta blocking strand has been displaced. We observe that this requires
$\approx 10^3\tau$ and appears to be largely independent of the length of
the toehold and adjacent domain. This is expected since this is only determined
by the time it takes the blue strand to diffuse and bind to the template when
the second toehold is exposed. 
The first transition is the displacement of the magenta blocking strand by the green
signal strand. This transition takes significantly longer time than the first
transition but is comparable with the duration of the second transition.
We observe a very strong dependence on toehold length and a much weaker dependence on domain length.
Increasing the toehold length from three to seven increases the duration of the
transition by more than an order of magnitude. This is consistent with
the exponential slowing down of the displacement kinetics when the toehold
is elongated~\citep{Yurke:2003}: the
magenta strand is most likely to thermally melt of the second toehold when
the green input strand is completely hybridized with the template. The
energetic barrier of thermal melting is exponential in the length of the
double stranded domain. The weaker dependence on domain length that we observe
is again due to the diffusion of the branching point when the magenta blocking
strand and the green input strand are competing for the template.
Finally, we observe that the total transition time is approximately
$2\times10^4\tau$. Remarkably, the total gate computation transition time
appears to be rather insensitive to the size of the toeholds and adjacent
domains. The distribution of gate operation times can be approximated by
the convolution of the three distributions associated with the three transitions
required for the gate operation. This suggests that the exponential toehold
dependence of the first transition is in part countered by the toehold size
independent second and third transitions. With the few data points we have
for the total gate operation time, it is difficult to extract any general
trends from our data.

We can characterize the gate fidelity as the probability that the red output
signal has been released after the gate transition. In the ideal case, this
probability would reproduce the logic table of an AND gate. We estimate the
fidelity by sampling conformations for $t\geq 8\times10^4$, where the AND
gate transition should have taken place. We have performed at least ten
statistical independent simulations of each AND gate which allows us to
estimate the accuracy of the fidelity. Tab.~\ref{tab:fidelity} shows the
fidelity of the AND gate for various simulated domain lengths. We
observe excellent fidelity for all logic states except the 01 input state
This shows that gate operation has completed in the time window we use to
estimate the fidelity.

In the 01 input state, only the blue input strand is present. Its low fidelity could be
attributed either to thermal melting of the red output strand or displacement of
the red output strand by the invading blue input strand due to transient defects.
If thermal melting was the dominant cause, we would expect the fidelity of the
10 and 00 logic states to be affected too. Hence we attribute the observed lack
of fidelity to defect induced strand displacement, where a transient bubble
allows the blue input strand to displace the red output signal. The binding
energy of this faulty state corresponds to that of the correct state, and hence
they appear equally likely (disregarding entropic contributions to the free energy).
This is in agreement with the fidelity $\approx 0.5$ that we observe in all
the simulations. Both the correct and the faulty states are long lived, since
they require the fortuitous opening of a bubble while the competing strand is
within hybridization distance. This explains why the accuracy for the 01 logic
state is much less than the other states, because different simulations
essentially sample only the correct or the faulty state. In experiments,
this might be suppressed by decreasing the concentration of strands, however to
study this further is outside the scope of the present paper.

\begin{table}[h]
\centering
\begin{tabular}{c|cccc}
$t$\textbackslash $l$ & $8$ & $10$ & $12$  \\ \hline
$3$ & & \begin{tabular}{|c|c|}
\noalign{\smallskip}\hline
11 &  $0.97 \pm 0.02$  \\
10 &  $0.00 \pm 0.00$  \\
01 &  $0.60 \pm 0.11$  \\
00 &  $0.03 \pm 0.02$  \\
\hline\noalign{\smallskip}
\end{tabular}
& \begin{tabular}{|c|c|}
\noalign{\smallskip}\hline
11 &  $0.90 \pm 0.06$  \\
10 &  $0.00 \pm 0.00$  \\
01 &  $0.37 \pm 0.15$  \\
00 &  $0.00 \pm 0.00$  \\
\hline\noalign{\smallskip}
\end{tabular}
\\
$5$ & & \begin{tabular}{|c|c|}
\noalign{\smallskip}\hline
11 &  $0.95 \pm 0.04$  \\
10 &  $0.05 \pm 0.03$  \\
01 &  $0.41 \pm 0.14$  \\
00 &  $0.01 \pm 0.01$  \\
\hline\noalign{\smallskip}
\end{tabular}
& \\
$7$ & \begin{tabular}{|c|c|}
\noalign{\smallskip}\hline
11 &  $0.90 \pm 0.06$  \\
10 &  $0.01 \pm 0.01$  \\
01 &  $0.65 \pm 0.14$  \\
00 &  $0.06 \pm 0.03$  \\
\hline\noalign{\smallskip}
\end{tabular}
& \begin{tabular}{|c|c|}
\noalign{\smallskip}\hline
11 &  $0.95 \pm 0.05$  \\
10 &  $0.02 \pm 0.01$  \\
01 &  $0.58 \pm 0.15$  \\
00 &  $0.00 \pm 0.00$  \\
\hline\noalign{\smallskip}
\end{tabular}
& \\
\end{tabular}
\label{tab:fidelity}
\caption{
Fidelity of AND gates for the simulated choices of toehold $t$ and adjacent $l$ domain sizes
for the four logic possibilities where both input strands are present (denoted 11), where
only the green input strand is present (denoted 10), where only the blue input strand is
present (denoted 01), and finally where none of the input strands are present (denoted 00).
The error estimate is obtained from the analysis of at least ten statistically independent
simulations.
}

\end{table}

\section{Conclusions\label{sec:Conclusions}}

We have studied strand displacement dynamics, in particular kinetics and
fidelity of an AND gate implemented using DNA strand displacement dynamics,
which form the basis of state-of-the-art DNA computing approaches.
The ideal causal strand displacement dynamics shown in the textbooks is one
where the first signal strand displaces a blocking strand to expose a second
toehold, which is hybridized with a second signal strand to finally release
the output strand indicating the true state of the DNA AND gate. Furthermore,
for the ideal AND gate, no signal is released if none or just one of the two
signal strands are present. However, chemical implementations of strand displacement
gates have to content with non-ideal behaviour that originates from thermal effects, 
which changes the textbook picture.

We have implemented a DNA AND gate within the framework of molecular dynamics
simulations utilizing a coarse-grained dynamic bonded molecular model of DNA
molecules that we have recently developed. We study an AND gate design
where a template strand has a first toehold, a first domain, a second toehold,
and a second domain. Initially the first toehold is exposed, while a blocking
strands is hybridized to the first domain and the second toehold. The output
strand is hybridized with the second domain only. The first input signal
hybridizes with the first toehold and partially displaces the blocking strand.
When the blocking strand dehybridises, the second toehold is exposed.
The second input strand can hybridise with the second toehold and displace
the output signal, hence completing the operation of the gate. We have
performed extensive simulations of this design to study gate operation for
toehold lengths from $3$ to $7$ and domains $8$ to $12$ nucleotides long for
systems of fixed temperature. Mapping these length ranges to melting
temperatures of DNA strands produces a range in excess of $200K$ from the
shortest toehold to the longest strand. Each simulation was performed with at
least ten statistical independent initial states.

In our analysis of the simulation results we observe a complex sequence of
dynamic transitions, where strands are released and rehybridize numerous times;
we see release of the output signal before either of the two input strands are
bound to the gate; and we even observe the release of the output signal in the absence of input
strands. All these effects are due to thermal melting and bubbles and are expected to
occur in a chemical implementations of strand displacement gates. We have
characterized the type of defects present in a DNA AND gate in the final state.
This shows that transient internal bubbles occurs with a high frequency, while
defects longer than 3 nucleotides are prevalently located at the end of the
template or at nicks between hybridized strands.

We have characterised the gate kinetics by analysing the distribution of times
it takes from hybridization to release of a strand, as well as the times required
from the release to rehybridization of a strand. These give insights into the
binding kinetics and bulk diffusion, respectively. 

We have also analysed the total gate operation time as well as the three transitions
that it is based on: hybridization of the first input signal, release of the
blocking strand, subsequent hybridization of the second input strand, and finally
release of the output signal. We see that the first displacement transition is
exponentially dependent on the toehold size, while the second and third
transitions are essentially independent of the toehold size.
We also see a weaker dependence on domain length of the displacement processes.
Remarkably the total operation time appears to be largely independent of both
the toehold and adjacent domain size at least within the range of lengths
studied in the present paper. This latter time distribution is approximately
the convolution of the time distributions associated with the three transitions, and
the second and third transitions effectively cancel the exponential toehold dependence
of the first transition. 

From the simulation results we can evaluate the fidelity of the gate operation.
Ideally, the probability of the output strand being released after the gate
has finished its operation should correspond to the logic table of an AND
operation. We observe excellent gate fidelity except for the input state
where only the second input signal is present, which has quite poor fidelity.
In this case, we observe that bubbles create transient toeholds that allow
the second strand to displace the output signal, and compete with the blocking
strand for the second toehold. This faulty state has the same number of
hybridization bonds and appears to be as favourable as the correct state where
the output signal remains hybridized with the template.

In conclusion, we have shown that computer simulations of the kind presented expose
the detailed molecular mechanisms behind non-ideality, and in particular how
non-ideality influences kinetics and fidelity of DNA strand displacement
computation. The presented simulation studies also iluminate how to
realise optimal implementations of logic gates in DNA strand displacement
operations, in terms of balancing high fidelity and fast computation transition times. 
We could also use simulations such as these as a starting point for developing and testing 
first principle statistical mechanical theories predicting the physical-chemical behaviour of DNA 
strand displacement systems.

\section{Acknowledgements}

The research leading to these results was sponsored in part by the 
European Community's Seventh Framework Programme under
grant agreement no 249032 (MATCHIT), in part by the Danish National 
Research Foundation and in part by the University of Southern Denmark. 
The work was conducted at the Center for Fundamental Living Technology 
(FLinT). We are grateful to Pierre-Alain Monnard for critical advice
concerning DNA properties.

\bibliographystyle{spbasic}

\end{document}